\documentstyle[aps,aps,epsf]{revtex}

\begin{document}
\draft
\twocolumn[\hsize\textwidth\columnwidth\hsize\csname 
@twocolumnfalse\endcsname
\title{The precision of slow-roll predictions for the CMBR anisotropies}
\author{J\'er\^ome Martin}
\address{DARC, Observatoire de Paris, \\ 
UMR 8629 CNRS, 92195 Meudon Cedex, France. \\
e-mail: martin@edelweiss.obspm.fr}
\author{Dominik J. Schwarz}
\address{Institut f\"ur Theoretische Physik, \\
Wiedner Hauptstra\ss e 8 -- 10, 1040 Wien, Austria. \\
e-mail: dschwarz@hep.itp.tuwien.ac.at}
\date{May 19, 2000}
\maketitle

\begin{abstract}
Inflationary predictions for the anisotropy of the cosmic microwave 
background radiation (CMBR) are often based on the slow-roll 
approximation. We study the precision with which the multipole moments of the 
temperature two-point correlation function can be predicted by means of 
the slow-roll approximation. We ask whether this precision is good enough 
for the forthcoming high precision observations by means of the MAP 
and Planck satellites. The error in the multipole moments due to the 
slow-roll approximation is demonstrated to be bigger than the error in the 
power spectrum. For power-law inflation with $n_{\rm S}=0.9$ the error from 
the leading order slow-roll approximation is $\approx 5\%$ for the amplitudes 
and $\approx 20\%$ for the quadrupoles. For the next-to-leading order the 
errors are within a few percent. The errors increase with 
$\vert n_{\rm S}-1\vert$. To obtain a precision of $1 \%$ it is necessary, but 
in general not sufficient, to use the next-to-leading order. 
In the case of power-law inflation this precision is obtained for 
the spectral indices if $\vert n_{\rm S}-1\vert < 0.02$ and for the 
quadrupoles if $\vert n_{\rm S}-1\vert < 0.15$ only. The errors in the higher 
multipoles are even larger than those for the quadrupole, e.g.~$\approx 15\%$ 
for $l=100$, with $n_{\rm S} = 0.9$ at the next-to-leading order. We 
find that the accuracy of the slow-roll approximation may be improved  
by shifting the pivot scale of the primordial spectrum (the scale 
at which the slow-roll parameters are fixed) into the regime of acoustic 
oscillations. Nevertheless, the slow-roll approximation cannot be improved 
beyond the next-to-leading order in the slow-roll parameters. 
\end{abstract}
\vspace*{1cm}
]

\section{Introduction}

High quality measurements of the Cosmic Microwave Background Radiation 
(CMBR) anisotropies have been published recently by the balloon borne 
experiments BOOMERanG \cite{Boomerang} and MAXIMA-1 \cite{Maxima}. A large 
number of multipoles ($26 \leq l \leq 625$ for BOOMERanG and 
$36 \leq l \leq 785$ for MAXIMA-1) has been covered by both experiments.
During the next years, high precision measurements will be performed 
by the MAP and Planck satellites \cite{MAP+Planck}. 
Inflation \cite{inflation} provides a mechanism to produce the primordial 
fluctuations of space-time and matter \cite{tensor,scalar,GP,MFB}, 
which lead to the CMBR anisotropies and to the large scale structure. 
This mechanism rests on the principles of General Relativity and Quantum 
Field Theory. It thus can be expected to get a hand on the physics 
of the very early Universe with 
help of the upcoming high precision measurements.
\par
The CMBR anisotropies are most conveniently expressed by the multipole 
moments $C_l$. The computation of the multipole moments requires the 
knowledge of the primordial spectrum and the transfer functions. The latter 
depend on the cosmological parameters $H_0, \Omega_{\rm M}, \Omega_{\Lambda}, 
\dots$. The transfer function characterizes 
the evolution of cosmological perturbations during the radiation and 
matter epochs. The primordial spectrum is predicted by inflation and depends 
on the evolution of the long wavelength perturbations during inflation and 
reheating. It can be predicted from a given model of inflation. 
\par
In this article, we will restrict our considerations to slow-roll 
inflation with one scalar field. This represents only a first step 
towards a more general study. Our aim is to address the following 
problems: What is the precision 
of the predicted multipole moments from the slow-roll approximation? 
Is this precision sufficient to reach the level of accuracy 
expected from the planned observations? Can the slow-roll approximation
be improved to arbitrary precision? 
\par
So far, the precision of the predicted power spectrum has been examined 
by Grivell and Liddle \cite{Grivell}. However, the power spectrum is 
not directly observable whereas the $C_l$'s are. We show 
that the error from the slow-roll approximation is important in the 
multipole moments. It is bigger than the error in the power spectrum. 
It turns out that the next-to-leading order slow-roll approximation \cite{SL}
is compulsory, but it may not be sufficient to reach an accuracy of a few \%
or less. 
\par
Wang, Mukhanov and Steinhardt \cite{WMS} have shown that predictions 
based on the time delay argument \cite{GP} or on the horizon crossing/Bessel
function approach (i.e.\ the slow-roll approximation) 
\cite{SL,Lea} are not reliable for general models of inflation. There have 
been various attempts to improve the slow-roll approximation to 
higher orders, see e.g.~\cite{SL,Coea,KV,Lea}. 
The conclusion of Wang et al. \cite{WMS} has been contested by 
Copeland et al.~\cite{Cea2}: ``We \dots conclude that any theoretical 
errors from the use of the slow-roll equations are likely to be subdominant''.
We show in this work that this claim is not correct unless the slow-roll 
parameters are extremely small. Typically, we find that the slow-roll 
parameters must be less than $0.01$ in order for the next-to-leading 
order to reach the level of precision of MAP or Planck. This means that 
there are models where the slow-roll error is dominant and 
the slow-roll approximation is valid. We find in agreement with 
the analysis in \cite{WMS} that a slow-roll approximation that goes beyond the 
next-to-leading order cannot exist. All higher order corrections  
are thus meaningless. In the derivation of this result we close a gap 
in the proof of the next-to-leading order equations. For some reason this 
gap was not noticed before in the literature. For this purpose we use and
generalize a new family of exact solutions, which was recently found by 
Starobinsky \cite{StaroExact}. 
\par
The scope of this paper is to quantify the error from the slow-roll
approximation. We compute the scalar multipole moments 
and the ratio $R\equiv C_2^{\rm T}/C_2^{\rm S}$ for 
power-law inflation for which the exact result is known. Then, we calculate 
the same quantities for the same model but in the context of the slow-roll 
approximation. The comparison of the two results provides an 
estimate of the error made by using the slow-roll approximation. We do not 
convolute this error with the uncertainties in the transfer functions. For 
the sake of clarity and simplicity we only make use of the transfer functions 
in the long wavelength limit. This approximation only mildly affects
the estimates of the error in the multipole moments. Then, we compare the 
slow-roll errors at leading and next-to-leading order to the cosmic variance.
Binning several multipoles together allows to reduce the cosmic variance, 
but does not reduce the slow-roll error. We find that the slow-roll error 
is hidden in the cosmic variance only for very small values of the slow-roll 
parameters $(< 10^{-2})$. We propose to reduce the slow-roll 
error by optimizing the pivot scale (the scale at which the slow-roll 
parameters are fixed) of the spectrum. However, this method is not 
sufficient to hide entirely the slow-roll error in the cosmic variance.
\par
This article is organized as follows: in the next section, the theory of 
cosmological perturbations and the calculations 
of the CMBR anisotropies are reviewed. Then, the low-$l$ multipole 
moments are computed exactly for power-law inflation (Sec.~III)
and approximately for slow-roll inflation (Sec.~IV). 
Comparison of the two results allows us to test the precision of the 
CMBR multipoles obtained from the slow-roll approximation in 
the last section. The slow-roll errors are shown to be observationally 
significant by comparing them with the cosmic variance. 
We set $c=\hbar = 1$ throughout the paper.

\section{From quantum fluctuations to CMBR anisotropies}

The line element for the spatially flat Friedmann-Lemaitre-Robertson-Walker 
background plus perturbations can be written as \cite{MFB}:
\begin{eqnarray}
\label{metricgi}
{\rm d}s^2 &=& 
a^2(\eta )\{-(1-2\phi ){\rm d}\eta ^2+2({\rm \partial}_iB){\rm d}x^i
{\rm d}\eta +[(1-2\psi )\delta _{ij} \nonumber \\
& &+2{\rm \partial }_i{\rm \partial }_jE+h_{ij}]{\rm d}x^i{\rm d}x^j\}\ .
\end{eqnarray}
In this equation, the functions $\phi $, $B$, $\psi $ and $E$ represent the 
scalar sector whereas the tensor $h_{ij}$, satisfying 
$h_i{}^i=h_{ij}{}^{,j}=0$, represents the gravitational waves. There are 
no vector perturbations because a single scalar field cannot 
seed rotational perturbations. The conformal time $\eta $ is related to the 
cosmic time $t$ by ${\rm d}t=a(\eta ){\rm d}\eta $. It is convenient 
to introduce the background quantity $\gamma (\eta )$ defined by $\gamma 
\equiv -\dot{H}/H^2$, where a dot means differentiation with respect to 
cosmic time and $H$ is the Hubble rate, $H\equiv \dot{a}/a$. Using 
conformal time we may write $\gamma =1-{\cal H}'/{\cal H}^2$, where 
${\cal H}\equiv a'/a$, and a prime denotes differentiation with 
respect to the conformal time. 
\par 
We assume that inflation is driven by a single scalar field. 
For the perturbations we introduce gauge-invariant variables \cite{B,MFB}, 
which reduce the equations of motion, in the small scale limit, 
to equations of harmonic oscillators \cite{Grigw,M,Griden,MS,MFB}. 
In the tensor sector (which is gauge invariant) 
we define the quantity $\mu _{\rm T}$ for each mode $k$ according to  
$h_{ij}=(\mu _{\rm T}/a)Q_{ij}(k)$, where $Q_{ij}(k)$ are the 
(transverse and traceless) eigentensors of the Laplace operator
on the spacelike sections and $k^2$ is the corresponding eigenvalue.
Gravitational waves do not couple to scalar fields. Thus the equation of 
motion is given by \cite{Grigw}:
\begin{equation}
\label{eomtensor}
\mu _{{\rm T}}''+\biggl[k^2-\frac{a''}{a}\biggr]\mu _{{\rm T}}=0\ .
\end{equation} 
The scalar sector is gauge dependent and the scalar perturbations 
of the metric are coupled to the perturbations of the stress tensor 
describing the matter. Fluctuations in the stress tensor involve 
perturbations in the energy density, ${\rm \delta}\rho$, and in the four 
velocity, ${\rm \delta }u^{\mu }=(-\phi/a,v^i/a)$. We describe 
perturbations in the density contrast by the gauge invariant 
quantity $\delta \equiv {\rm \delta }\rho /\rho +(\rho '/\rho )(B-E')$. 
Perturbations in the velocity can be written as 
$v_i\equiv {\rm \partial }_iw+w_i$. Since we are interested in the 
scalar sector, only the first term has to be taken into account. We choose 
to work with the gauge invariant quantity $v\equiv w+E'$. Scalar 
perturbations of the geometry can be characterized by the two gauge 
invariant Bardeen potentials $\Phi Q\equiv \phi +(1/a)[(B-E')a]'$ and 
$\Psi Q\equiv \psi -{\cal H}(B-E')$ \cite{B},where $Q(k)$ is a scalar 
harmonic. During inflation, the Universe is dominated by the scalar 
field $\varphi=\varphi _0(\eta )+\varphi _1(\eta )Q$. Fluctuations in 
the scalar field are characterized by the gauge invariant 
quantity ${\rm \delta}\varphi \equiv \varphi _1+\varphi _0'(B-E')$. In 
this simple case, the time evolution of fluctuations can be reduced 
to the study of the equation of motion for the variable 
$\mu _{\rm S}\equiv -\sqrt{2\kappa}\, a[{\rm \delta }\varphi +
(\varphi _0'/{\cal H})\Phi]$, where $\kappa \equiv 8\pi G$. Its 
equation of motion is very similar to that of the gravitational 
waves \cite{M,Griden}:
\begin{equation}
\label{eomscalar}
\mu _{{\rm S}}''+\biggl[k^2-
\frac{(a\sqrt{\gamma})''}{(a\sqrt{\gamma })}\biggr]\mu _{{\rm S}}=0\ .
\end{equation}

The integration of (\ref{eomtensor}) and (\ref{eomscalar})
leads to the primordial spectrum of the fluctuations. 
For the initial conditions we assume that the scalar and tensor 
perturbations are in the quantum vacuum state when the scale of interest 
was well within the Hubble radius ($1/k_{\rm ph} \ll c/H$) during the 
early stages of inflation. Therefore all fluctuation variables are quantum 
operators during inflation. After inflation, the Universe 
is filled with baryons, photons, 
neutrinos and (cold) dark matter. For that epoch, the perturbed Einstein 
equations cannot be reduced to the simple form of Eqs.~(\ref{eomtensor})
and (\ref{eomscalar}) and need to be integrated numerically. This leads to 
the transfer functions. 
\par
The cosmological perturbations induce anisotropies in the temperature of the 
CMBR, which have been detected by COBE \cite{COBE1} first. This is the
Sachs-Wolfe effect \cite{SW}. Since it does not depend on the photon 
frequency, the black body shape of the photon spectrum is conserved from 
the last scattering surface to its observation today \cite{FIRAS}.  
The measured anisotropies in the photon intensity translate into 
anisotropies in the temperature of the black body. 

For the temperature fluctuations we introduce the abbreviation 
$\Delta (\vec{e})\equiv ({\rm \delta }T/T)(\vec{e})$, where $\vec{e}$ 
characterizes the direction of the beam on the celestial sphere. The 
contributions of the scalar and tensor perturbations are given by:
\begin{eqnarray}
\label{ds}
\Delta^{\rm S}(\vec{e}) &=&
\frac14 \delta_{\gamma} + \Phi - e^i 
{\rm \partial }_i v 
+ \int_{\eta_{\rm lss}}^{\eta_0} {\rm d} \bar{\eta }
{\partial\over \partial \bar{\eta}} (\Phi + \Psi)\ ,\\
\label{dt}
\Delta^{\rm T} (\vec{e}) &=&
- \frac12 e^i e^j \int_{\eta_{\rm lss}}^{\eta_0} 
{\rm d} \bar{\eta } {\partial\over \partial \bar{\eta }} h_{ij}\ . 
\end{eqnarray}
The first three terms of the scalar contribution are evaluated on the 
last scattering surface, i.e.~at $\eta_{\rm lss}$. They represent the intrinsic 
fluctuations, the Sachs-Wolfe effect and the Doppler effect. The forth term
is the so-called integrated Sachs-Wolfe effect. The integration is performed
along the photon trajectory, which is parameterized by the conformal time here.
$\eta_0$ denotes the conformal time at observation today. $\delta _{\gamma }$
is the perturbed density contrast of the photons and $v$ the perturbed 
velocity of the photon fluid. For large angular scales only the first 
two terms are important. For isentropic (sometimes called adiabatic) 
perturbations the scalar part reduces to:
\begin{equation}
\label{largescale}
\Delta^{\rm S}(\vec{e}) = \frac13 \Phi(\vec{e}) +(...)\ .
\end{equation}
Usually, the CMBR anisotropies are expressed through the multipole 
moments $C_l$. The $C_l$ are the coefficients in an expansion over 
Legendre polynomials of the CMBR temperature two-point correlation: 
\begin{equation}
\langle \Delta^{\rm S,T}(\vec{e}_1) \Delta^{\rm S,T}(\vec{e}_2)\rangle = 
\frac 1{4\pi} \sum_l (2l+1) C_l^{\rm S,T} P_l(\cos\delta) \ , 
\end{equation}
where $\cos\delta \equiv \vec{e}_1 \cdot \vec{e}_2$. The brackets 
$\langle \rangle$ denote the averaging over many ensembles. Averages over 
many ensembles cannot be replaced by spatial averages on the celestial sphere 
due to the lack of ergodicity of the stochastic process $\Delta (\vec{e})$, 
see Ref. \cite{GM}. If, nevertheless, we do this the error made can be 
quantified by means of the cosmic variance.
\par
The computation of the multipoles for a given model requires the 
knowledge of the initial spectrum of the fluctuations and of the transfer 
function. The power spectrum of the Bardeen potential is defined 
in terms of the two-point correlator for the operator 
$\hat{\Phi }(\eta ,{\bf x})$:
\begin{equation}
\label{defPPhi}
\langle 0|\hat{\Phi }(\eta ,{\bf x})\hat{\Phi }(\eta ,{\bf x}+{\bf r})|0\rangle
\equiv 
\int _0 ^{\infty } \frac{{\rm d}k}{k}\frac{\sin kr}{kr}k^3P_{\Phi }(\eta ,k)\, .
\end{equation}
Similarly, the power spectrum of gravitational waves is defined as:
\begin{equation}
\label{defPgw}
\langle 0|\hat{h}_{ij}(\eta ,{\bf x})\hat{h}^{ij}(\eta ,{\bf x}+{\bf r})
|0\rangle 
\equiv \int _0 ^{\infty } \frac{{\rm d}k}{k}\frac{\sin kr}{kr}k^3P_{h }
(\eta ,k)\, .
\end{equation}
A priori, the primordial power spectra are time dependent quantities. However, 
for the multipoles between $l=2$ and $l=2000$, we are interested in scales 
which are well beyond the horizon at the end of inflation. In a first 
approximation for those scales the power spectra do not evolve in time 
during inflation and they can be written as: 
\begin{eqnarray}
\label{Pasymp}
k^3 P_\Phi(k) &=& A_{\rm S}^i(k_0)\left(\frac{k}{k_0}\right)^{n_{\rm S}-1}, \\
\label{Pasymp2}
\ k^3 P_h(k) &=& A_{\rm T}^i(k_0)\left(\frac{k}{k_0}\right)^{n_{\rm T}},
\end{eqnarray}
where the spectral indices $n_{\rm S}$, $n_{\rm T}$ and the amplitudes 
$A_{\rm S}^i$, $A_{\rm T}^i$ are independent quantities and $k_0$ is 
an arbitrarily fixed scale which is introduced to link various 
notations in the literature. The spectral indices can also be determined 
from $n_{\rm S}-1\equiv {\rm d}\ln(k^3P_{\Phi })/{\rm d}\ln k$ 
and $n_{\rm T}\equiv {\rm d}\ln(k^3P_{h })/{\rm d}\ln k$.
\par
An accurate calculation of the multipole moments requires numerical 
computations. However, for small $l$, the approximate equation 
(\ref{largescale}) can be used. For density perturbations \cite{P82} this 
leads to: 
\begin{eqnarray}
\label{cls}
C_l^{\rm S} &=& \frac{4\pi}{9} \int _0^\infty {{\rm d} k\over k}
j_l^2(k r_{\rm lss}) 
T_{\Phi }(k r_{\rm lss} \rightarrow 0) \nonumber \\
& & \times A_{\rm S}^i 
\left(k\over k_0\right)^{n_{\rm S} -1} , 
\end{eqnarray}
where $j_l$ is the spherical Bessel function of order $l$ and 
$r^{\rm ph}_{\rm lss} \equiv a(\eta_0) r_{\rm lss} = 
a_0 (\eta_0 - \eta _{\rm lss}) \approx a_0 \eta_0$ $\approx 2 R_{\rm H}$ is 
the comoving line-of-sight distance to the last scattering surface. 
\par
$T_{\Phi }(k r_{\rm lss} \rightarrow 0)$ is approximately the transfer 
function for superhorizon modes. It is $k$ independent and 
therefore only the amplitude is modified but not the spectral index. 
The domain of validity of the latter approximation can be evaluated 
as follows. In the integral (\ref{cls}) the main contribution comes 
from the modes around $k r_{\rm lss}\sim l+1$. We use that to estimate 
for which multipole moments the $k$ independent transfer function is 
good enough. The mode whose wavelength is equal to the Hubble radius today, 
i.e.~such that $2\pi a_0/k=l_H(\eta_0)$, has 
$k\eta_{\rm 0} = 4\pi$. Therefore the constant superhorizon transfer function 
is a reasonably good approximation if $(l+1)/r_{\rm lss}\ll 
4\pi/\eta_0$, that is to say $l \ll 10$. This is a very optimistic estimate 
since it does not take into account the first approximation, made in 
Eq.~(\ref{largescale}), on which the validity of Eq.~(\ref{cls}) rests. 
\par
With the above approximations the low-$l$ multipoles can be calculated exactly 
\cite{P82}. The result reads:
\begin{eqnarray}
\label{C2s}
C_l^{\rm S} &=& \pi^{3/2} {\Gamma[(3 - n_{\rm S})/2]
\Gamma[l+(n_{\rm S}-1)/2]\over \Gamma[(4-n_{\rm S})/2] 
\Gamma[l+2-(n_{\rm S}-1)/2]} \nonumber \\
& & \times (k_0 r_{\rm lss})^{1 - n_{\rm S}}
{A_{\rm S}\over 9} \ ,
\end{eqnarray}
where $A_{\rm S}\equiv 
A_{\rm S}^iT_{\Phi }(k r_{\rm lss}\rightarrow 0)$. For gravitational 
waves we obtain the following expression \cite{Staro}:
\begin{eqnarray}
\label{clgw}
C_l^{\rm T} &=& \frac{9 \pi}{4} (l-1)l(l+1)(l+2)
\left(k_0 r_{\rm lss}\right)^{- n_{\rm T}} \nonumber \\
& &\times \int_0^\infty {{\rm d} y\over y}
\left| I_l(y)\right|^2 A_{\rm T}y^{n_{\rm T}},
\end{eqnarray}
where the function $I_l(y)$, $y \equiv k r_{\rm lss}$, is defined by:
\begin{equation}
\label{Il}
I_l(y)\equiv  \int_{0}^{y} {j_2(x)j_l(y - x)\over x (y-x)^2} {\rm d} x\ .
\end{equation}
The superhorizon transfer function for gravitational waves does not 
appear explicitly because it is equal to one. As a consequence we can 
write $A_{\rm T}^i\equiv A_{\rm T}$. The computation of $C_l^{\rm T}$ is 
more complicated than the calculation of $C_l^{\rm S}$. The integral 
$I_l$ can be calculated exactly in terms of 
special functions, see Ref.~\cite{Staro}. However, the second integration 
over $k$ cannot be performed analytically and we must rely on numerical 
integration. 
\par
Below we will be interested in the ratio of tensor to scalar quadrupole 
contributions \cite{Dea,SS,Cea,St}: 
\begin{equation}
\label{defR}
R \equiv {C_2^{\rm T}\over C_2^{\rm S}}\ .
\end{equation}
Expressed in terms of the tensor spectral index this is the so-called 
consistency equation of inflation. 
\par
We have seen that the calculation of $C_l^{\rm S}$ requires the knowledge of 
the transfer function and of the primordial spectrum. In principle, 
$T_{\Phi }(kr_{\rm lss})$ is known accurately as the result of numerical 
calculations, e.g.\ see Refs.~\cite{transfer}. When we calculate 
the multipoles using Eqs.~(\ref{cls}) or (\ref{clgw}) we make two
approximations: a long wavelength approximation for the transfer function 
and we neglect the contribution of radiation (pure matter assumption) to 
the expansion of the Universe at photon decoupling. The long wavelength 
approximation results in neglecting other contributions besides 
(\ref{largescale}) in the Sachs-Wolfe effect for scalars and in considering 
that the tensor and scalar superhorizon transfer functions are constant. 
The pure matter assumption results in a small error in the numerical value of 
$T_{\Phi}(kr_{\rm lss} \to 0)$. However, for small values of $l$ these 
errors are small. 

\begin{figure}
\begin{center}
\leavevmode
\hbox{%
\epsfxsize=8.5cm
\epsffile{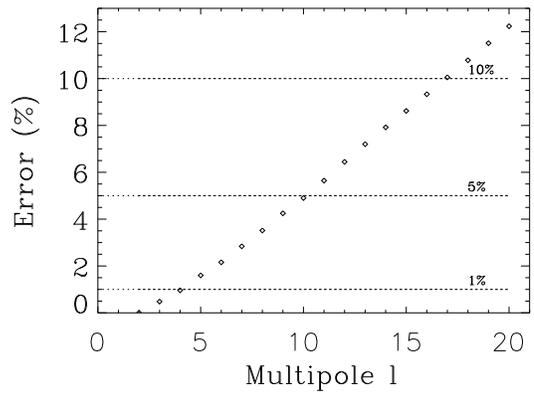}
}
\end{center}
\caption{Error due to the long wavelength approximation in the transfer 
function for the scalar multipoles with a flat primordial spectrum. The 
exact multipoles are calculated by means of the CMBFAST code and are 
normalized to the quadrupole.}
\label{errcmb}
\end{figure}

In order to test this claim quantitatively and to quantify the 
contribution to the total error coming from the 
transfer function (see Fig.~\ref{errcmb}), we
compute the scalar multipole moments for low $l$ numerically with 
CMBFAST \cite{SZ} for the following values of the cosmological 
parameters: $H_0=50 \mbox{km/s/Mpc}, \Omega_0 = 1, \Omega_{\rm CDM} = 0.95, 
\Omega_{\rm B} = 0.05$. We compare them to the multipole moments given by 
Eq.~(\ref{cls}) with a constant transfer function. The code CMBFAST 
automatically normalizes to the COBE result \cite{COBE1}. The result 
is expressed by the band powers $(\delta T_l/T_0)^2 \equiv l(l+1)C_l/(2\pi)$, 
where $T_0 \approx 2.73$ K is the average temperature of the CMBR. 
For a flat ($n_{\rm S} =1$) primordial spectrum, CMBFAST gives 
$\delta T_2 \approx 27.5 \mu K$ or $Q_{\rm rms-PS}\approx 17.8 \mu K$,
where the quadrupole rms fluctuation is given by $Q_{\rm rms-PS} \equiv T_0 
\sqrt{(5/4\pi)C_2}$. We normalize the amplitude $A_{\rm S}$ in 
Eq.~(\ref{cls}) is to the latter value of the 
quadrupole. In Fig.~\ref{errcmb} we plot the differences of both 
calculations, divided
by the CMBFAST results, and express this number as the error in \%.
The error in the quadrupole (\ref{C2s}) vanishes ``by construction''. 
Equation (\ref{cls}) shows that ${\rm \delta }T_l= T_0 \sqrt{A_{\rm S}}/3 
(k_0 r_{\rm lss})^{(1 - n_{\rm S})/2} = 
Q_{\rm rms-PS}\sqrt{12/5}$, whereas the CMBFAST-$\delta T_l$ is $l$-dependent,
despite both band powers are calculated from the same primordial spectrum. 
The difference between both band powers is exclusively due to the use of 
different transfer functions and to the neglection of the Doppler  
and integrated Sachs-Wolfe effects. In this way, we can isolate and 
estimate the error coming from the long wavelength approximation, being 
given that the spectrum is normalized to COBE. 

A similar study has been done in Ref.~\cite{BN}. The errors 
given in that article differ from those obtained here because a 
different normalization is used. In Ref.~\cite{BN}, the spectrum 
is normalized to the multipole moment $C_{10}$ instead of the 
quadrupole. As a consequence, in that case the error in $C_{10}$ vanishes 
``by construction''. 

Figure \ref{errcmb} confirms the importance of the transfer function and the 
analytical estimates made at the beginning of this article. The error 
is below $1\%$ only for $l<4 $. For $C_{10}$, which is often 
used to normalize the spectrum, the effect of the subleading terms
in $k$ is already $5\%$. The error from the pure matter assumption has 
not been fully accounted for by this method, because we do not test the
error in the numerical value of $T_{\Phi}(kr_{\rm lss} \to 0)$ when we 
normalize the quadrupole to the COBE result. Since this error is a pure 
overall numerical factor, it does not affect our conclusions.

\section{Predictions of power-law inflation}

In this section, we turn to the study of power-law inflation. This model is 
of particular importance because it allows to calculate all quantities of 
interest exactly. Moreover, this exact result is at the basis of the 
slow-roll approximation. 
\par
Power-law inflation is given by the following solution for the scale 
factor and the scalar field:
\begin{equation}
\label{defpl}
a(\eta )=l_0|\eta |^{1+\beta }, \quad \varphi =\varphi_{\rm i} +
\frac{m_{\rm Pl}}{2}\sqrt{\frac{\gamma  }{\pi }}(1+\beta )\ln |\eta |,
\end{equation}
where $m_{\rm Pl}$ is the Planck mass and $\varphi_{\rm i}$ is the initial 
value of the scalar field at conformal time $\eta_{\rm i}$. In this model 
inflation occurs if $\beta < -2$ (we do not consider the case where 
$-2 < \beta < -1$ which cannot be realized with a single scalar field). 
The quantity $l_0$ has the dimension of a length and its value will roughly 
determine the amplitude of the CMBR fluctuations today. In the particular 
case of power-law inflation, the function $\gamma(\eta)$ is a constant equal 
to $(2+\beta )/(1+\beta)$. For $-\infty < \beta <-2$, $\gamma$ goes from one 
to zero, this last value corresponding to the de Sitter spacetime. The scale 
factor and scalar field of Eqs.~(\ref{defpl}) are solutions of the Einstein 
equations for the scalar field potential:
\begin{equation}
\label{pot}
V(\varphi)= V_{\rm i} \exp\left[\frac{4\sqrt{\pi }}{m_{\rm Pl}}\sqrt{\gamma}
(\varphi - \varphi_{\rm i})\right]\ ,
\end{equation}
where $V_{\rm i}$ is the value of the potential at $\eta_{\rm i}$. 

\subsection{Density perturbations}

The effective potential for density perturbations, $U_{\rm S}\equiv 
(a\sqrt{\gamma })''/(a\sqrt{\gamma })$, see Eq.~(\ref{eomscalar}), reads:
\begin{equation}
\label{potU}
U_{\rm S}(\eta) = {(\beta+1)\beta\over \eta^2} \ .
\end{equation}
This simple form of the potential allows an exact integration of 
Eq.~(\ref{eomscalar}). The solution is expressed in terms of Bessel 
functions. This provides the initial power spectrum, i.e.~$A_{\rm S}^i$ and 
$n_{\rm S}$. In order to evolve the superhorizon spectrum, we can rely on 
the conservation law \cite{L,MS} for the quantity: 
$\zeta \equiv ({\cal H}^{-1}\Phi '+\Phi)/\gamma +\Phi$. This gives the 
superhorizon transfer function: $T_{\Phi }(k r_{\rm lss} \rightarrow 0)=
[9(2\beta +3)^2]/[25\gamma ^2(1+\beta )^2]$. Then the amplitude of the scalar 
quadrupole and the spectral index take the form:
\begin{equation}
\label{Anpl}
A_{\rm S}(k_0)=\frac{l_{\rm Pl}^2}{l_0^2}\frac{9}{25\pi \gamma }f(\beta ) 
k_0^{n_{\rm S} -1} \ , 
\quad n_{\rm S}=
2\beta +5=\frac{1-3\gamma  }{1-\gamma } \ ,
\end{equation}
where 
\begin{equation}
\label{deff}
f(\beta) \equiv 
\frac{1}{\pi }\biggl[\frac{\Gamma(-\beta -1/2)}{2^{\beta +1}}\biggr]^2\ ,
\end{equation}
which is unity for $\beta =-2$. As expected, the amplitude of scalar 
perturbations is roughly determined by the ratio $l_{\rm Pl}/l_0$. Very 
often the final spectrum is expressed in terms of the Hubble rate at some 
time $\eta_*$, instead of the scale $l_0$. We have $H_* \equiv H(\eta_*) = 
- [(1+ \beta)/l_0] |\eta_*|^{-2-\beta}$. Therefore the amplitude reads:
\begin{equation}
\label{AsplH}
A_{\rm S}(k_0)=l_{\rm Pl}^2 H_*^2 \frac{9}{25\pi\gamma} f(\beta) 
|1+\beta|^{2(\beta +1)}\biggl(\frac{k_0}{a_*H_*}\biggr)^{n_{\rm S}-1}\ .
\end{equation}
The amplitude $A_{\rm S}$ is displayed as a function of $\gamma$ in 
Fig.~\ref{Apl}. It diverges in the de Sitter limit $\gamma \to 0$.
COBE measured the spectral index to be $n_{\rm S} = 1.2\pm 0.3$ \cite{COBE1}. 
The $1(2) \sigma$ value $n_{\rm S} = 0.9(0.6)$ corresponds to 
$\gamma \approx 0.048(0.167)$.

\subsection{Gravitational waves}

\begin{figure}
\setlength{\unitlength}{\linewidth}
\begin{picture}(1,0.6)
\put(0.54,0.05){\makebox(0,0){$\gamma$}}
\put(0.03,0.35){%
\makebox(0,0)[b]{\shortstack{\scriptsize $
\frac{A_{\rm S, T}}{(l_{\rm Pl} H_*)^2}
\times (\frac{k_0}{a_*H_*})^{1-n_{\rm S}}$}}%
}%
\put(0.54,0.53){\makebox(0,0){Power-law inflation}}
\put(0.49,0.33){\makebox(0,0){\epsfig{figure=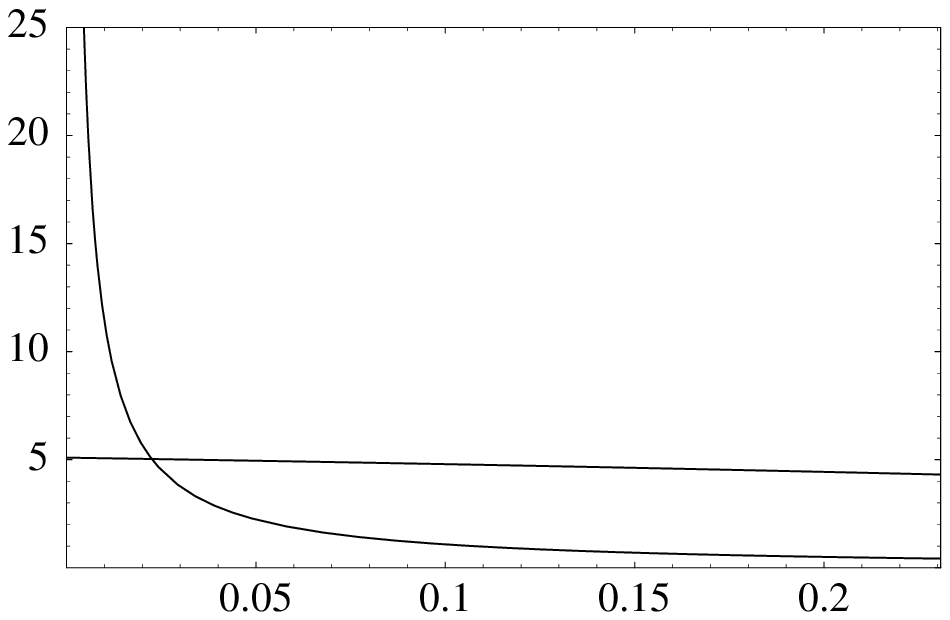,width=0.8\linewidth}}}
\end{picture}
\caption{
\label{Apl}
The amplitudes of scalar and tensor perturbations. In the de Sitter
limit $\gamma \to 0$ the scalar amplitude diverges. For larger values 
of $\gamma$ the perturbations are dominated by the tensor mode.}
\end{figure}

The calculation of the spectrum for gravitational waves is performed along 
the same lines as above. The effective potential is the same as 
for density perturbations, i.e. $U_{\rm T}\equiv a''/a=
\beta (1+\beta )/\eta ^2$. Since the superhorizon transfer function is equal 
to one, $A_{\rm T}$ and $n_{\rm T}$ can be written as:
\begin{equation}
\label{Atntpl}
A_{\rm T}(k_0)=\frac{l_{\rm Pl}^2}{l_0^2}\frac{16}{\pi }f(\beta ) k_0^{n_{\rm T}}, 
\quad n_{\rm T}=2\beta +4=-\frac{2\gamma }{1-\gamma  } \ .
\end{equation}
For power-law inflation, the relation $n_{\rm S}=n_{\rm T}+1$ holds. 
In terms of $H_*$ the amplitude is given by 
\begin{equation}
\label{AtplH}
A_{\rm T}(k_0)= l_{\rm Pl}^2 H_*^2 \frac{16}{\pi }f(\beta)
|1+\beta|^{2(\beta +1)}\biggl(\frac{k_0}{a_*H_*}\biggr)^{n_{\rm T}}\ .
\end{equation}
Figure \ref{Apl} shows the scalar and tensor amplitudes (\ref{AsplH}) and
(\ref{AtplH}), respectively. For $\gamma > 9/400 = 0.0225$ the tensor mode 
dominates.

\subsection{Multipole moments}

\begin{figure}
\setlength{\unitlength}{\linewidth}
\begin{picture}(1,0.6)
\put(0.54,0.05){\makebox(0,0){$\gamma$}}
\put(0.03,0.35){
\makebox(0,0)[b]{\shortstack{\scriptsize $
\frac{C_2^{\rm S, T}}{(l_{\rm Pl} H_*)^2}\times 
(r_{\rm lss}a_*H_*)^{n_{\rm S}-1}$}}%
}
\put(0.54,0.53){\makebox(0,0){Power-law inflation}}
\put(0.49,0.33){\makebox(0,0){\epsfig{figure=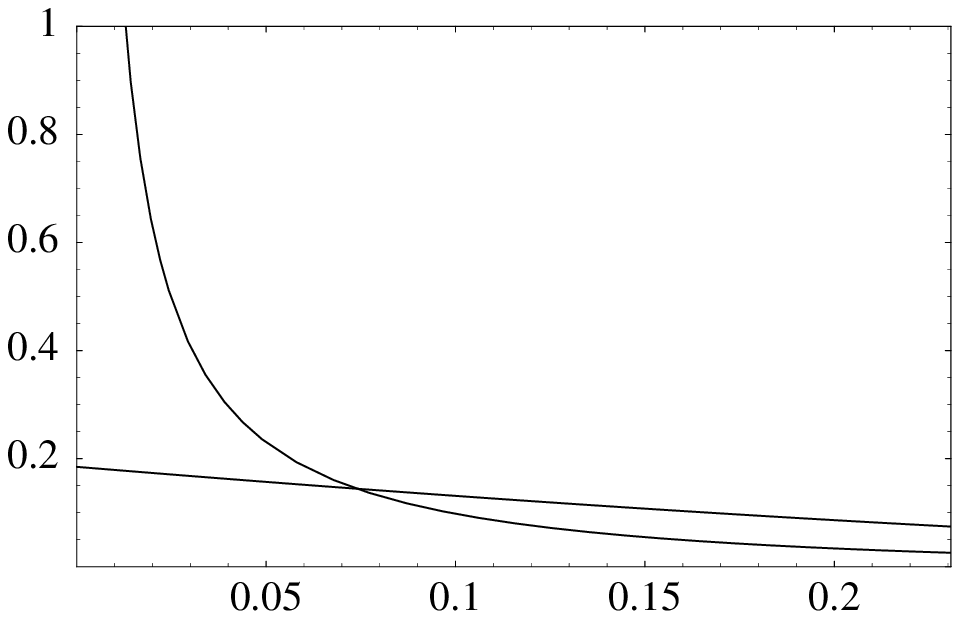,width=0.8\linewidth}}}
\end{picture}
\caption{
\label{Cpl}
The quadrupole moments of scalar and tensor perturbations.}
\end{figure}

The multipole moments predicted by power-law inflation can easily be 
computed from Eqs.~(\ref{C2s}) and (\ref{clgw}). The quadrupoles are 
displayed in Fig.~\ref{Cpl}. Compared to the amplitudes the 
importance of the tensor mode is slightly suppressed, it becomes 
the dominant mode at $\gamma \gtrsim 0.07$, which corresponds to 
$n_{\rm S} \lesssim 0.85$.

\begin{figure}
\setlength{\unitlength}{\linewidth}
\begin{picture}(1,0.6)
\put(0.54,0.05){\makebox(0,0){$\gamma$}}
\put(0.05,0.35){\makebox(0,0){$R$}}
\put(0.54,0.53){\makebox(0,0){Power-law inflation}}
\put(0.49,0.33){\makebox(0,0){\epsfig{figure=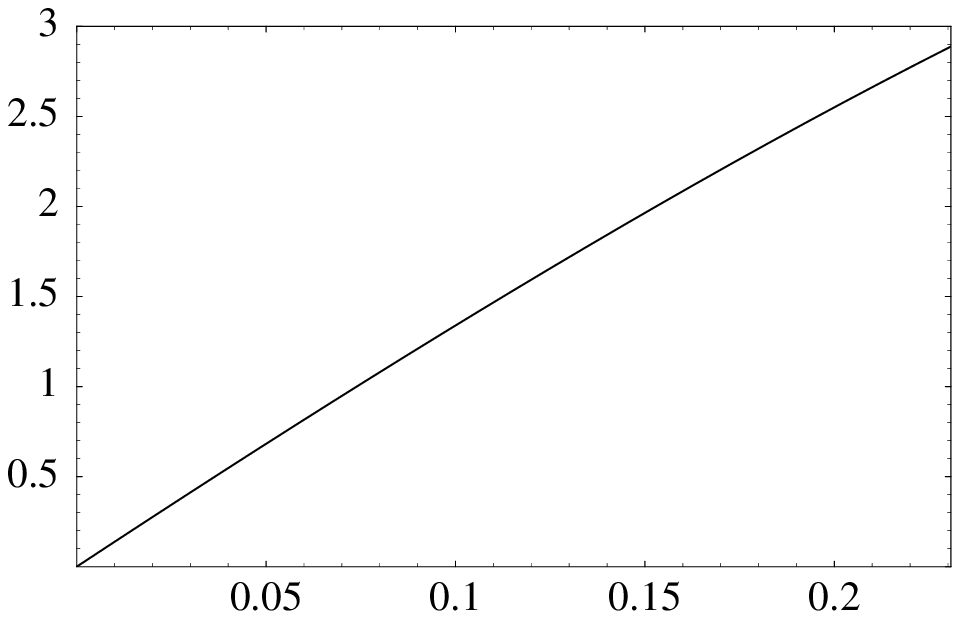,width=0.8\linewidth}}}
\end{picture}
\caption{
\label{Rpl}
The tensor to scalar ratio of the quadrupole moments.}
\end{figure}

We calculate the ratio $R$ for power-law inflation: 
\begin{equation}
\label{ccpl}
R= 13.86 \gamma F[n_{\rm T}(\gamma)] =
-6.93\frac{n_{\rm T}}{1-\frac{n_{\rm T}}{2}}F(n_{\rm T})\ ,
\end{equation}
where the function $F(n_{\rm T})$ is given by: 
\begin{eqnarray}
\label{defF}
F(n_{\rm T})& \equiv & 496.1 \times 2^{1-n_{\rm T}} 
\frac{\Gamma^2(\frac{3-n_{\rm T}}{2})\Gamma(4-\frac{n_{\rm T}}{2})}{
\Gamma (2-n_{\rm T})\Gamma (2+\frac{n_{\rm T}}{2})} \nonumber \\
& & \times \int _0^{\infty }{\rm d}kk^{n_{\rm T}-1}|I_2(k)|^2. 
\end{eqnarray}
In this expression we have used the equation $n_{\rm S}=n_{\rm T}+1$, valid 
for power-law inflation only, to express everything in terms of 
$n_{\rm T}$. We have $\int _0^{\infty }{\rm d}kk^{-1}|I_2(k)|^2= 2.139 \times
10^{-4}$ such that $F(n_{\rm T}=0)=1$. Notice that the factors 
$k_0r_{\rm lss}$ and $k_0/(a_*H_*)$ cancel in $R$ 
because $n_{\rm S}=n_{\rm T}+1$. $R$ versus $\gamma $ is 
plotted in Fig.~\ref{Rpl}. This plot demonstrates that within the $2\sigma$ 
error bars of COBE, there is a large parameter space where the tensor 
mode dominates the scalar modes, see e.g.~Refs.~\cite{SS,St} for a more 
detailed discussion. 

\section{Predictions of slow-roll inflation}

\begin{figure}
\begin{center}
\leavevmode
\hbox{
\epsfxsize=8.5cm
\epsffile{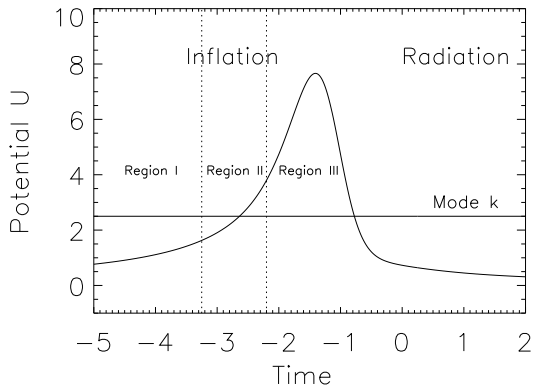}}
\end{center}
\caption{Sketch of the effective potential for density perturbations and/or 
gravitational waves during inflation and radiation.}
\label{potentialdp}
\end{figure}

For a general model of inflation exact solutions are not available. 
Generically, the potentials $U_{\rm S}$ and $U_{\rm T}$ are different 
but nevertheless their shape is similar. A sketch of the generic form of
$U_{\rm S}$ and $U_{\rm T}$ is displayed in Fig.~\ref{potentialdp}.
The details of the realistic reheating transition are not taken into 
account in this simple figure. During the radiation dominated era 
the potential goes to zero, since $a \propto \eta$. 

For a given mode $k$, the inflationary epoch can be divided into three 
stages, see Fig.~\ref{potentialdp}. In region I the mode $k$ is 
subhorizon. In that case the effective 
potential is small compared to $k^2$. In the limit $k/(aH) \to \infty$ for 
fixed $k$, the vacuum fluctuations are given by, see Ref.~\cite{MS}:
\begin{equation}
\label{regionI} 
\mu_{\rm S,T}(\eta) \rightarrow \mp 4 \sqrt{\pi} l_{\rm Pl} 
\frac{e^{- i k(\eta -\eta_{\rm i})}}{\sqrt{2 k}}\ , 
\end{equation}
respectively. In region III the mode is superhorizon. In the 
limit $k/(aH) \to 0$ at fixed $k$, the potential term is dominant, 
and the ``exact'' solutions read: 
\begin{eqnarray}
\mu _{\rm S}(\eta) &=& C_{\rm S}(a\sqrt{\gamma})(\eta) \times \nonumber \\
\label{regionIII} 
& & \left[1 - k^2\int^\eta{1\over (a^2 \gamma)(\bar{\eta})}\int^{\bar{\eta}} 
(a^2\gamma)(\tilde{\eta})\,{\rm d}\tilde{\eta}\,{\rm d}\bar{\eta}\right]\  ,\\
\mu _{\rm T}(\eta) &=& C_{\rm T}a(\eta)\ . 
\end{eqnarray}
Usually, density perturbations are described in terms of the Bardeen 
potential $\Phi$ instead in terms of $\mu _{\rm S}$. 
The order $k^2$ term is necessary to obtain the leading order expression 
for the Bardeen potential, since 
$\Phi =[{\cal H}\gamma /(2k^2)][\mu _{\rm S}/(a\sqrt{\gamma })]'$, 
see Refs. \cite{MS,Griden}. Thus, in region III, the superhorizon Bardeen 
potential is given by:
\begin{equation}
\label{BardeenIII}
\Phi (\eta)=
\frac{C_{\rm S}{\cal H}}{2a^2}\int ^{\eta }a^2\gamma\,{\rm d}\bar{\eta }\ .
\end{equation}
Our aim is to calculate the spectra at the end of inflation, i.e.~in region 
III. The time dependence of the solutions in this region is known and 
the difficulty lies in the calculation of the constants $C_{\rm S}$ and 
$C_{\rm T}$. Since the solutions are uniquely determined in region I, 
this amounts to join the super- and subhorizon solutions. Therefore we need 
to know the behavior of the perturbations in region II. 

A popular approach is the slow-roll approximation \cite{SL,Lea}. 
The idea is that there was an epoch during inflation where the scalar 
field was rolling down its potential $V(\varphi)$ very slowly. Under certain 
conditions (see below) this is close to the behavior during power-law inflation
and the exact solutions from power-law inflation are used in region II 
to interpolate between the sub- and superhorizon solutions.
\par
Slow roll is controlled by the three (leading) slow-roll parameters 
(see e.g.~Ref.~\cite{Lea}) defined by: 
\begin{eqnarray}
\label{defepsilon}
\epsilon &\equiv & 3 \frac{\dot{\varphi}^2}{2} 
\left(\frac{\dot{\varphi}^2}{2} + V\right)^{-1} = -\frac{\dot{H}}{H^2}\ , \\
\label{defdelta}
\delta &\equiv & -\frac{\ddot{\varphi }}{H\dot{\varphi }} =
- \frac{\dot{\epsilon }}{2 H \epsilon }+\epsilon\ , \\
\label{defxi}
\xi &\equiv & \frac{\dot{\epsilon }-\dot{\delta }}{H}\ .
\end{eqnarray}
We see in particular that $\gamma(\eta)=\epsilon$ in region II. The 
equations of motion for $\epsilon$ and $\delta$ can be written as:
\begin{equation}
\label{eqmotionsrpara}
\frac{\dot{\epsilon}}{H}=2\epsilon (\epsilon -\delta)\ , \quad 
\frac{\dot{\delta }}{H}=2\epsilon (\epsilon -\delta )-\xi\ .
\end{equation}
The slow-roll conditions are satisfied if $\epsilon$ and $\delta$ are much 
smaller than one and if $\xi = {\cal O}(\epsilon^2,\delta^2,\epsilon\delta)$. 
{}From Eqs. (\ref{eqmotionsrpara}), it is clear that this amounts 
to consider $\epsilon$ and $\delta$ as constants. This property is 
crucial for the calculation of the perturbations. 
\par
For power-law inflation the slow-roll parameters satisfy:
\begin{equation}
\label{srparapl}
\epsilon = \delta < 1\ ,\quad \xi = 0 \ .
\end{equation}
Therefore the slow-roll conditions are fulfilled if   
$\epsilon \ll 1$, that is to say if $\beta $ is close to $-2$ (scale 
invariance). In fact, the slow-roll approximation is an expansion 
around power-law inflation with $0 < -(\beta +2) \ll 1$. To illustrate 
this point, let us consider the exact equation:
\begin{equation}
\label{etasr}
\eta = - \frac{1}{aH} + \int{\rm d}a \frac{\epsilon}{a^2H} \ .
\end{equation}
If we assume that $\epsilon$ is a constant, the previous equation reduces 
to $aH \approx -(1+\epsilon)/\eta$. This is equivalent to a scale 
factor which behaves like:
\begin{equation}
\label{srscalefactor}
a(\eta)\approx l_0 |\eta|^{-1-\epsilon}.
\end{equation}
Interestingly enough, the effective power index at leading order depends 
on $\epsilon$ only. 

\subsection{Density perturbations}

The effective potential of density perturbations can be calculated in terms 
of the slow-roll parameters exactly. The result is:
\begin{equation}
\label{potsr}
U_{\rm S}(\eta) = a^2H^2 [2 - \epsilon + (\epsilon -\delta)(3-\delta )+\xi] \ .
\end{equation}
In the slow-roll approximation $a^2 H^2 \approx \eta^{-2}(1+ 2\epsilon)$ and
the effective potential reduces to 
$U_{\rm S}\approx (2+6\epsilon -3\delta )\eta ^{-2}$. Since $\epsilon$ and 
$\delta$ must be seen as constants in the slow-roll approximation, the 
equation of motion (\ref{eomscalar}) is of the same type as in  
power-law inflation. The solutions are given by Bessel functions:
\begin{equation}
\label{regionII}
\mu _{\rm S}=(k\eta )^{1/2}[B_1J_{\nu _{\rm S}^{\rm (sr)}}(k\eta )
+B_2J_{-\nu _{\rm S}^{\rm (sr)}}(k\eta )],
\end{equation}
whose order is given by 
\begin{equation}
\nu _{\rm S}^{\rm (sr)} = -\frac{3}{2}-2\epsilon +\delta \ . 
\end{equation}
A comment is in order here: The potential $U_{\rm S}$ depends on the 
scale factor and its derivatives only. One could think, looking at 
Eq.~(\ref{srscalefactor}), that $U_{\rm S}$ also depends on $\epsilon$ only. 
This is not the case. The reason is that $U_{\rm S}$ contains terms 
like $\dot{\epsilon }/\epsilon $ (for example) which are linear 
in $\delta$, see Eqs.~(\ref{eqmotionsrpara}). First one must 
calculate all derivatives, replace them with their expression in terms of 
$\epsilon$ and $\delta$, and only then consider that the slow-roll 
parameters are constant.
\par
We would also like to stress that keeping higher orders in $\epsilon$ does 
not make sense. If terms of quadratic order in the slow-roll parameters 
are kept, the solution for density perturbations in region II can no 
longer be expressed in terms of Bessel functions. This is because the 
slow-roll parameters can no longer be considered as constant in 
time, see Eqs. (\ref{eqmotionsrpara}). Therefore any considerations 
at this order in the framework of the slow-roll approximation is 
meaningless. The same conclusion has been obtained by Wang, Mukhanov, and 
Steinhardt \cite{WMS}.
\par
Let us now calculate the constant $C_{\rm S}$. The first step is to match 
the solutions of region I and II. This procedure fixes $B_1$ and $B_2$. 
Using Eqs.~(\ref{regionI}) and (\ref{regionIII}), one obtains: 
$B_1/B_2 = - e^{i\pi \nu_{\rm S}^{\rm (sr)}}$ and 
$B_1=2i\pi l_{\rm Pl}\exp [i\nu_{\rm S}^{\rm (sr)}(\pi/2)-i(\pi /4)+
ik\eta_{\rm i}]/(\sqrt{k}\sin\pi\nu_{\rm S}^{\rm (sr)})$. 
Note that $B_1$ and $B_2$ do not 
depend on the time at which the matching between regions I and II 
is performed. The joining between regions II and III remains to 
be performed at some time $\eta _{\rm S}$, which will be fixed below. 
Expanding everything up to next-to-leading order in the slow-roll parameters, 
one obtains:
\begin{eqnarray}
\label{C_S}
|C_{\rm S}|^2=\frac{l_{\rm Pl}^2}{l_0^2}\frac{8\pi }{\epsilon }
\biggl[1 &-& 2(C+\ln k)(2\epsilon -\delta ) \nonumber \\
&+& 2(\delta -\epsilon)\ln |\eta _{\rm S}|\biggr]k^{-3},
\end{eqnarray}
with $C\equiv \gamma_E + \ln 2 - 2 \approx - 0.7296$, 
$\gamma _{\rm E} \approx 0.5772$ being the Euler constant. 
The Bardeen potential given in Eq.~(\ref{BardeenIII}) is now completely 
specified. Note that $\gamma(\eta)$ in Eq.~(\ref{BardeenIII}) 
is a time dependent function, evaluated in region III, whereas $\epsilon$ in
Eq.~(\ref{C_S}) is a constant parameter, which is fixed by 
$\epsilon = \gamma(\eta _{\rm S})$.
\par
For scalar perturbations it is useful to evaluate the quantity
\begin{equation}
\zeta = -{\mu_{\rm S}\over 2a \sqrt{\epsilon}} \ ,
\end{equation}
which is a {\em constant} for the dominant mode at superhorizon 
scales \cite{L,MFB,MS}. The quantity $-\zeta $ is denoted ${\cal R}$ in 
Ref.~\cite{Lea}. Instead of expressing the spectrum in terms 
of the ratio $l_{\rm Pl}/l_0$, it is usual to write it in terms 
of the Hubble rate at some time $\eta _*$. Of course, there is 
nothing deep in this choice and one could have kept working 
with $l_{\rm Pl}/l_0$. A priori, the value of $\eta _*$ is arbitrary 
and could either be in regions I, II or III. However, in order to make contact 
with the literature, we will assume that $\eta _*$ is in region II. Then, in 
the slow-roll approximation, the value of $H(\eta _*)$ can be written as: 
\begin{equation}
\label{H}
H_* \equiv H(\eta _*)=\frac{1}{l_0}\left[1+\epsilon (1+\ln |\eta _*|)\right]\ .
\end{equation} 
In Ref. \cite{Lea}, $\eta _*$ is the time which satisfies the 
relation $a(\eta _*)H(\eta _*)=k$ for each mode $k$. 
In other words, we have $\eta _*=\eta _*(k)$. In this article, we adopt another
convention and choose $\eta _*$ such that it is not a function 
of $k$. Then, a straightforward calculation gives 
\begin{eqnarray}
\label{zetasr}
k^3 P_{\zeta}(k) &=& \frac{l_{\rm Pl}^2H^2_*}{\pi \epsilon }
\biggl\{1-2\epsilon -  2\left[C + \ln k|\eta _*| \right](2\epsilon -\delta) +
\nonumber \\ 
& & 2(\delta -\epsilon)
\ln\biggl\vert\frac{\eta _{\rm S}}{\eta_*}\biggr\vert\biggr\}\ .
\end{eqnarray}
\par
The matching time $\eta_{\rm S}$ remains to be fixed by a physical 
argument. To our knowledge, this issue has been overlooked in the literature 
so far. All works on the slow-roll approximation, starting with 
Ref.~\cite{SL}, have tacitly assumed that $\eta_{\rm S}/\eta_* = 1$, 
without further justification. A priori, an equally good choice would be, for 
example, when the mode $a_*H_*$ crosses the effective potential, i.e. when 
$(1+2\epsilon )/\eta _*^2=U_{\rm S}(\eta_{\rm S})$. It is easy to show 
that this boils down 
to the choice $\eta_{\rm S}/\eta _*=\sqrt{2}$. It is important to realize 
that different choices for the ratio $\eta_{\rm S}/\eta_* $ lead to 
different observational predictions. Although a change in $\eta_{\rm S}$ 
would not
change the spectral index, it would change the amplitude of scalar 
perturbations and the ratio of tensor to scalar contributions $R$.
\par 
The missing physical argument comes from a new family of exact solutions 
which has a slow-roll regime in a certain limit. One exact solution is 
of course power-law inflation, but it does not help for the purpose of fixing
$\eta_{\rm S}/\eta_*$, because the spectrum does not depend on 
$\eta_{\rm S}/\eta_*$ for $\delta = \epsilon$. These solutions are found 
by the ansatz 
\begin{equation}
\label{ansatz}
a \sqrt{\gamma} = {A \over |\eta|^\alpha} \ ,
\end{equation}
where $A$ and $\alpha $ are two free parameters. This defines a two-parameters
family of exact solution. Note that this family is not equivalent 
to power-law inflation. The power-law model 
($A = l_0 \sqrt{\gamma(\beta)}$, $\alpha = - 1 - \beta$) is a subclass of 
this two-parameter family. Of course, this is because 
$a(\eta )\propto |\eta |^{- \alpha }$ is just a solution of 
Eq.~(\ref{ansatz}), 
viewed as a second order differential equation for the scale 
factor, but not the general solution. The limit $A$ to zero and $\alpha $ close 
to one gives a slow-roll inflation model. The particular case $\alpha =1$ 
was already found recently by Starobinsky \cite{StaroExact}. This one 
parameter family of solutions is characterized by a flat spectrum, 
$n_{\rm S} = 1$. Eq.~(\ref{ansatz}) is a generalization of 
Starobinsky's original ansatz \cite{StaroExact}. The spectrum may be 
calculated exactly to read
\begin{equation}
\label{zetaexact}
k^3 P_{\zeta}(k) = {l_{\rm Pl}^2\over \pi ^2 A^2}  
2^{2\alpha} \Gamma ^2\biggl(\alpha + \frac{1}{2}\biggr)k^{-2(\alpha -1)} \ .
\end{equation}
The case $\alpha =1$ gives $k^3 P_{\zeta}(k)=l_{\rm Pl}^2/(\pi A^2)$ and 
coincides with the result of Ref.~\cite{StaroExact}.
\par
We now need to calculate the slow-roll spectrum for this new class of 
solutions. A comparison with Eq.~(\ref{zetasr}) will allow us to fix the 
ratio $\eta _{\rm S}/\eta _*$. Let us first determine the slow-roll 
parameters. In the slow-roll approximation we find
\begin{equation}
\label{srz'z}
{(a\sqrt{\gamma})'\over a\sqrt{\gamma}} = 
- {1\over\eta} (1 + 2\epsilon -\delta) \ ,
\end{equation}
whereas insertion of the ansatz (\ref{ansatz}) into this equation 
gives $(a\sqrt{\gamma})'/(a\sqrt{\gamma})=-\alpha /\eta $. Therefore, one 
has $2\epsilon =(\alpha -1)+\delta $ and especially $2 \epsilon=\delta $ if 
$\alpha =1$. It is interesting to note that we no longer have the 
relation $\epsilon =\delta$ typical of power-law inflation. Let us also 
emphasize that the two-parameter family is the only family of exact solutions 
which permits a slow-roll approximation. Eq.~(\ref{srz'z}) is a 
necessary condition for the validity of the slow-roll approximation. This 
equation can be viewed as a first-order differential equation for the 
quantity $a\sqrt{\gamma }$. Integration of this equation leads to the ansatz 
given in Eq. (\ref{ansatz}). Therefore our determination of the ratio 
$\eta _{\rm S}/\eta _*$ is general. The value of $A$ in the slow-roll 
limit is obtained from $A = a\sqrt{\gamma}|\eta|^\alpha$ and is expressed 
in terms of $H_*$ with the help of (\ref{etasr}). This gives 
$A^2 = \epsilon H_*^{-2} [1 + 2\epsilon + 
2(2\epsilon - \delta)\ln |\eta _*|]$. 
Thus we obtain the slow roll spectrum from Eq.~(\ref{zetaexact}):
\begin{eqnarray}
k^3 P_{\zeta}(k) &=& \frac{l_{\rm Pl}^2H^2_*}{\pi \epsilon}
\left[1-2\epsilon \right. \nonumber \\
& & - \left. 2 \left(C + \ln k|\eta _*|\right)
(2\epsilon -\delta)\right].
\label{zetasrf}
\end{eqnarray}
A comparison with (\ref{zetasr}) shows that 
\begin{equation}
\label{comp}
\eta_{\rm S} = \eta_*.
\end{equation}
Note that we could have derived the slow-roll spectrum of $\zeta$ from 
the exact spectrum (\ref{zetaexact}) right from the beginning, 
by approximating it in the slow-roll regime. However, we have chosen to 
take the Bessel function/horizon crossing approach, because it is this 
approach which has been discussed in the literature.
Let us note that the transfer function for $\zeta$ is unity. 
This means that the spectrum of $\zeta$ during the matter dominated era
is identical to the spectrum at the end of inflation (region III).
\par
We are mostly interested in the spectrum of the metric potential $\Phi$ since 
this quantity appears in the calculations of the multipole moments, see 
Eq.~(\ref{cls}). If we assume that the Universe is matter dominated at 
the surface of last scattering, then the conservation law provides 
us with the relation $\zeta = (5/3) \Phi$. Then, the spectrum of the 
Bardeen potential follows from (\ref{zetasrf}) as: 
\begin{eqnarray}
\label{nsrs}
n_{\rm S}^{({\rm sr})} &=& 1 - 4\epsilon + 2\delta, \\
\label{Asrs}
A_{\rm S}^{({\rm sr})} &=& \frac{9 l_{\rm Pl}^2 H^2_*}{25\pi \epsilon}  
\left[1 - 2\epsilon - 2(C+\ln k_0|\eta _*|)(2\epsilon - \delta)\right].
\end{eqnarray} 
These expressions are consistent with (4.3) and (5.1) of \cite{Lea}. 
The amplitude of scalar perturbations blows up when the slow-roll 
approximation becomes accurate, i.e.~when $\epsilon$ goes to zero. 
\par
To end this section, let us make a last comment. It is clear from the 
previous considerations that we need the slow-roll approximation 
in region II only. In particular, this scheme of approximation is not 
needed in region III since the ``exact'' solution is known. 
However, one may wish to use it in region III also. Then, in this region, 
the Bardeen potential is given by $\Phi \approx 
(C_{\rm S}/2)\epsilon (1-3\epsilon +2\delta )$. 
The long-wavelength transfer function, which allows to pass from the end 
of inflation to the matter dominated epoch can be expressed as 
$T_{\Phi }\approx [9/(25\epsilon ^2)](1+6\epsilon -4\delta )$. Using the 
two previous formula, one can show that one recovers the spectrum given 
in Eqs.~(\ref{nsrs}) and (\ref{Asrs}). However, in principle, this 
method is not appropriate since we use an approximated solution whereas 
an exact one is available.

\subsection{Gravitational waves}

For gravitational waves, the same lines of reasoning can be applied. In 
region II, the effective potential can be written as:
\begin{equation}
\label{potgw}
U_{\rm T}(\eta ) = a^2 H^2 \left(2 - \epsilon\right), 
\end{equation}
and gives in the slow-roll limit
\begin{equation}
\label{potgwsr}
U_{\rm T}(\eta ) \sim \frac{2 + 3\epsilon}{\eta ^2} \ .
\end{equation}
Therefore the matching of sub- and superhorizon solution is again 
reduced to power-law inflation. The solution of $\mu _{\rm T}$ is similar 
to the one given in Eq.~(\ref{regionII}), where the effective index of the 
Bessel function is now given by:
\begin{equation}
\label{betagwsr}
\nu _{\rm T}^{({\rm sr})} =-\frac{3}{2}-\epsilon\ .
\end{equation}
This solution can be used to find the constant $C_{\rm T}$. Then, the power 
spectrum of gravitational waves reads 
\begin{equation}
k^3 P_h(k) = {l^2_{\rm Pl}\over l^2_0}
{16\over \pi} \left(1 - 2 C \epsilon - 2 \epsilon\ln k \right) \ , 
\end{equation}
from which we deduce that:
\begin{eqnarray}
\label{nsrgw}
n_{\rm T}^{({\rm sr})} &=& -2\epsilon , \\ 
\label{Ansrgw}
A_{\rm T}^{({\rm sr})} &=& l^2_{\rm Pl} H^2_* {16\over \pi} 
\left[1 - 2 (C+ 1+\ln k_0|\eta _*|) \epsilon \right] \ .
\end{eqnarray} 
We see that there exists a crucial difference between density 
perturbations and gravitational waves. In the case of 
gravitational waves, the ambiguity related to the choice of the 
matching time is not present. 

\begin{figure}
\setlength{\unitlength}{\linewidth}
\begin{picture}(1,0.6)
\put(0.54,0.05){\makebox(0,0){$\epsilon$}}
\put(0.03,0.35){\makebox(0,0){$A_{\rm S, T}\over (l_{\rm Pl} H_*)^2$}}
\put(0.54,0.53){\makebox(0,0){Slow-roll inflation}}
\put(0.49,0.33){\makebox(0,0){\epsfig{figure=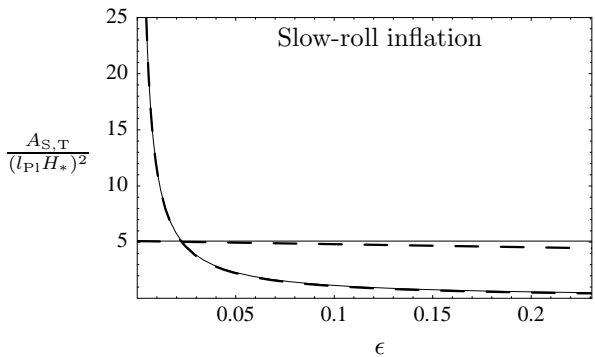,width=0.8\linewidth}}}
\end{picture}
\caption{
\label{srAa}
The scalar and tensor amplitudes from the slow-roll approximation for 
$\epsilon = \delta$. The scalar amplitude diverges in the de Sitter limit
$\epsilon \to 0$. The leading order is drawn by full lines, the  
next-to-leading order by dashed lines. We have set $k_0|\eta _*|=1$.}
\end{figure}

\begin{figure}
\setlength{\unitlength}{\linewidth}
\begin{picture}(1,0.6)
\put(0.54,0.05){\makebox(0,0){$\epsilon$}}
\put(0.03,0.35){\makebox(0,0){$A_{\rm S, T}\over (l_{\rm Pl} H_*)^2$}}
\put(0.54,0.53){\makebox(0,0){Slow-roll inflation}}
\put(0.49,0.33){\makebox(0,0){\epsfig{figure=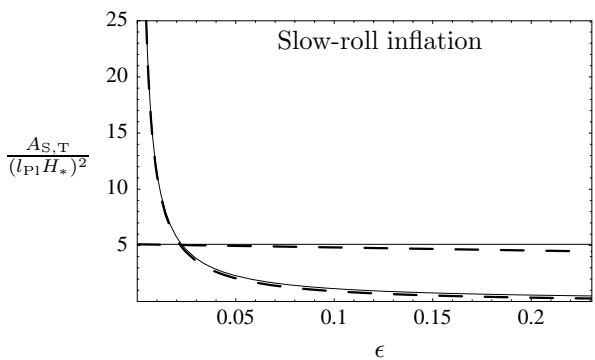,width=0.8\linewidth}}}
\end{picture}
\caption{
\label{srAb}
The same as Fig.~\ref{srAa}, but for $\delta = 2 \epsilon$.}
\end{figure}

The amplitudes of scalar and tensor modes versus the slow-roll 
parameter $\epsilon $ are displayed in Fig.~\ref{srAa} for $\delta = 
\epsilon$ and in Fig.~\ref{srAb} for $\delta = 2 \epsilon$
at leading and next-to-leading order. The first case is an approximation to 
the exact power-law result, the case $\delta = 2\epsilon$ is the slow-roll 
approximation to Starobinsky's exact solution.  

\subsection{Multipole moments}

Let us first start with the calculation of $C_l^{\rm S}$. We write the scalar 
multipoles as 
\begin{equation}
C_l^{\rm S} \equiv g_l(n_{\rm S}) \frac{A_{\rm S}}9 \ , 
\end{equation}
which defines the function $g_l(n_{\rm S})$, cf.~Eq.~(\ref{C2s}). 
To compute $C_l^{\rm S}$ at the next-to-leading order in the slow-roll 
parameters we need to expand $g_l(n_{\rm S})$ to first order around 
$n_{\rm S}=1$,
since $n_{\rm S}$ itself is a linear function of $\epsilon$ and $\delta$:
\begin{eqnarray}
\label{dlP1}
g_l(n_{\rm S}) &= & g_l(n_{\rm S}=1)+(n_{\rm S}-1)
\frac{{\rm d}g_l}{{\rm d}n_{\rm S}}(n_{\rm S}=1) \\
\label{dlP2}
&=& \frac{2\pi}{l(l+1)}[1+(n_{\rm S}-1)(D_l-\ln k_0r_{\rm lss})] \ ,
\end{eqnarray}
where
\begin{equation}
\label{Dl}
D_l \equiv  1 - \ln 2 + \Psi(l) + {l+1/2\over l(l+1)} 
\end{equation}
and $\Psi(x) \equiv {\rm d}\ln \Gamma (x)/{\rm d}x$.  
For the quadrupole we have $D_2 \approx 1.1463$ and for large $l$, 
$D_l = 1 + \ln(l/2)+ {\cal O}(1/l)$,
due to $\Psi(l) = \ln l + {\cal O}(1/l)$. 
Using Eqs.~(\ref{nsrs}) and (\ref{Asrs}), we find the scalar multipoles at 
next-to-leading order as
\begin{eqnarray}
\label{Clsr}
C_l^{\rm S} &=& \frac{2l_{\rm Pl}^2H_*^2}{25\epsilon}\frac{1}{l(l+1)}
\biggl[1-2\epsilon-2(D_l+C)(2\epsilon -\delta) \nonumber \\
& & +2(2\epsilon -\delta )\ln \frac{r_{\rm lss}}{|\eta _*|}\biggr] \ .
\end{eqnarray}
In this equation the Doppler effect, the integrated Sachs-Wolfe effect, 
and the evolution of the transfer function are neglected. In the next 
section we will argue that this does not prevent the estimation of the 
slow-roll error. 

We now calculate $R$ in the slow-roll regime at the leading and 
next-to-leading order. The scalar quadrupole follows from (\ref{Clsr})
and reads:
\begin{eqnarray}
\label{C2first}
C_2^{\rm S} &=& \frac{l_{\rm Pl}^2 H_*^2}{75\epsilon}
\biggl[1-2\epsilon-2(D_2+C)(2\epsilon -\delta) \nonumber \\
& & +2(2\epsilon -\delta )\ln \frac{r_{\rm lss}}{|\eta _*|}\biggr] \ ,
\end{eqnarray}
where $D_2+C\approx 0.4167 $.
\par
\begin{figure}
\setlength{\unitlength}{\linewidth}
\begin{picture}(1,0.6)
\put(0.54,0.05){\makebox(0,0){$\epsilon$}}
\put(0.03,0.35){\makebox(0,0){$C_2^{\rm S, T}\over (l_{\rm Pl} H_*)^2$}}
\put(0.54,0.53){\makebox(0,0){Slow-roll inflation}}
\put(0.49,0.33){\makebox(0,0){\epsfig{figure=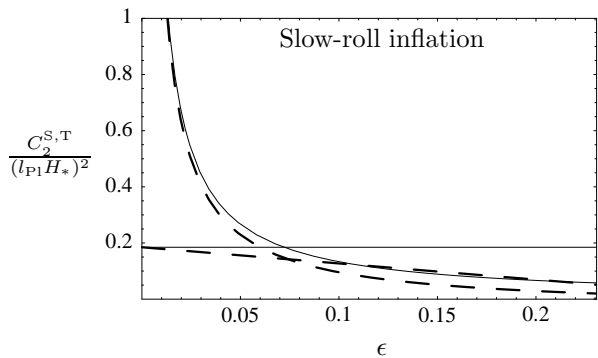,width=0.8\linewidth}}}
\end{picture}
\caption{
\label{sra}
The scalar and tensor quadrupole moments from the slow-roll approximation for
$\epsilon = \delta$. The leading order is drawn by full lines, the
next-to-leading order by dashed lines. We have set $r_{\rm lss}=|\eta _*|$.}
\end{figure}

Let us now compute $C_2^{\rm T}$. Using Eqs.~(\ref{clgw}) and (\ref{Ansrgw}), 
we find at next-to-leading order
\begin{equation}
\label{C2gwsr}
C_2^{\rm T}=0.1848 l_{\rm Pl}^2 H_*^2 \left[1-2\biggl(B+C+1-
\ln \frac{r_{\rm lss}}{|\eta _*|} \biggr)\epsilon\right] \ ,
\end{equation}
where the number B is defined by 
$B\equiv \int _0^{\infty }{\rm d}kk^{-1} \ln(k) I_2^2(k)/
\int _0^{\infty }{\rm d}kk^{-1} I_2^2(k) 
\approx 1.2878$ so that $B+C\approx 0.5582 $. For higher tensor multipoles 
the numerical values of 
the constants in (\ref{C2gwsr}) are modified, but not the functional 
dependence on the slow-roll parameters.
In Fig.~\ref{sra} the scalar and tensor quadrupoles at leading and 
next-to-leading orders are displayed for the case $\epsilon =\delta $. 

Taking into account the expressions for $C_2^{\rm S}$ and $C_2^{\rm T}$ given 
previously, we finally find the following expression for R in the slow-roll 
regime: 
\begin{equation}
\label{Rsr}
R=13.86\epsilon\left[1+0.5504\epsilon-0.8334\delta - 
2(\epsilon - \delta) \ln \frac{r_{\rm lss}}{|\eta _*|} \right] \ .
\end{equation}
At leading order we recover the so-called 
{\em consistency condition for slow-roll inflation} \cite{Lea}, which reads
\begin{equation}
\label{consistency}
R = - 6.93 n_{\rm T} \ .
\end{equation}
This equation cannot be generalized by the use of (\ref{Rsr}) to a 
next-to-leading order equation, because it would involve the knowledge of
the order ${\cal O}(\epsilon^2)$ terms in $n_{\rm T}$. As discussed above,
terms of that order are not meaningful in the slow-roll approximation.
\par
\begin{figure}
\setlength{\unitlength}{\linewidth}
\begin{picture}(1,0.6)
\put(0.54,0.05){\makebox(0,0){$\epsilon$}}
\put(0.05,0.35){\makebox(0,0){$R$}}
\put(0.54,0.53){\makebox(0,0){Slow-roll inflation}}
\put(0.49,0.33){\makebox(0,0){\epsfig{figure=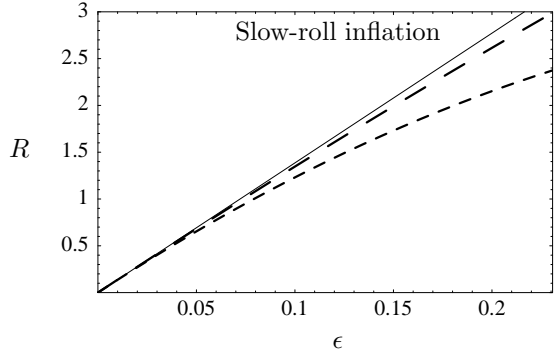,width=0.8\linewidth}}}
\end{picture}
\caption{
\label{srR}
The tensor to scalar ratio at leading order (full line) and at next-to-leading
order for $\delta = \epsilon$ (long dashed line) and $\delta = 2 \epsilon$
(short dashed line). The leading order is independent of 
$\delta$. We have set $r_{\rm lss}=|\eta _*|$.}
\end{figure}

In Fig.~\ref{srR}, the ratio $R$ is displayed at leading and 
next-to-leading order for the two cases $\epsilon =\delta $ and 
$2\epsilon =\delta $.

\section{Discussion of errors}

The aim of this section is to quantify the magnitude of the error introduced 
by the slow-roll approximation. For this purpose, we compare the slow-roll 
predictions with the exact results of power-law inflation. We explicitly  
test the following quantities: 
$Q\in \{n_{\rm S}-1,n_{\rm T},A_{\rm S},A_{\rm T},
C_l^{\rm S},C_2^{\rm T},R\}$, i.e.~quantities related to the 
power spectra and the quadrupole moments. 
\par
We denote by $Q$ the exact result of power-law inflation and 
by $Q^{(0)}$, $Q^{(1)}$ the slow-roll results at leading and 
next-to-leading orders, respectively. The error is 
estimated by calculating:
\begin{equation}
\label{deferr}
e_{Q}^{(i)}\equiv \biggl\vert \frac{Q^{(i)}-Q}{Q}\biggr\vert \times 100 \%\ .
\end{equation}
Let us start with an estimate of the errors in the prediction of the 
spectral indices $n_{\rm S}$ and $n_{\rm T}$. For the leading order 
slow-roll approximation $n_{\rm T} = n_{\rm S} - 1 = 0$ and thus the error
is $e^{(0)}_{n_{\rm S}-1}=e^{(0)}_{n_{\rm T}}=100\% $, except for 
de Sitter inflation. It is absolutely compulsory 
to use the next-to-leading order result for the spectral indices. We express 
the error as a function of $\gamma$. The best slow-roll approximation to a
power-law model is given by $\epsilon = \delta=\gamma$, and therefore 
$n_{\rm T}^{\rm (sr)} = n_{\rm S}^{\rm (sr)} - 1$ for this case. Thus from
Eqs.~(\ref{Anpl}), (\ref{Atntpl}), (\ref{nsrs}), and (\ref{nsrgw})  
the error in the spectral indices from the slow-roll approximation is
\begin{equation}
e^{(1)}_{n_{\rm S}-1}=e_{n_{\rm T}}^{(1)} = \gamma \times  100 \% \ .
\end{equation} 
Thus the next-to-leading 
order slow-roll approximation predicts the spectral indices
with an error less than $1 \%$, if $\gamma < 0.01$ or 
$0.98 < n_{\rm S} < 1$.

\begin{figure}
\setlength{\unitlength}{\linewidth}
\begin{picture}(1,0.6)
\put(0.51,0.05){\makebox(0,0){$\gamma=\epsilon=\delta$}}
\put(0.5,0.3){\makebox(0,0){$e_{A_{\rm S}}^{(0)}$}}
\put(0.73,0.21){\makebox(0,0){$e_{A_{\rm S}}^{(1)}$}}
\put(0.3,0.5){\makebox(0,0){$e_{C_2^{\rm S}}^{(0)}$}}
\put(0.75,0.45){\makebox(0,0){$e_{C_2^{\rm S}}^{(1)}$}}
\put(0.6,0.53){\makebox(0,0){Error in \% }}
\put(0.45,0.33){\makebox(0,0){\epsfig{figure=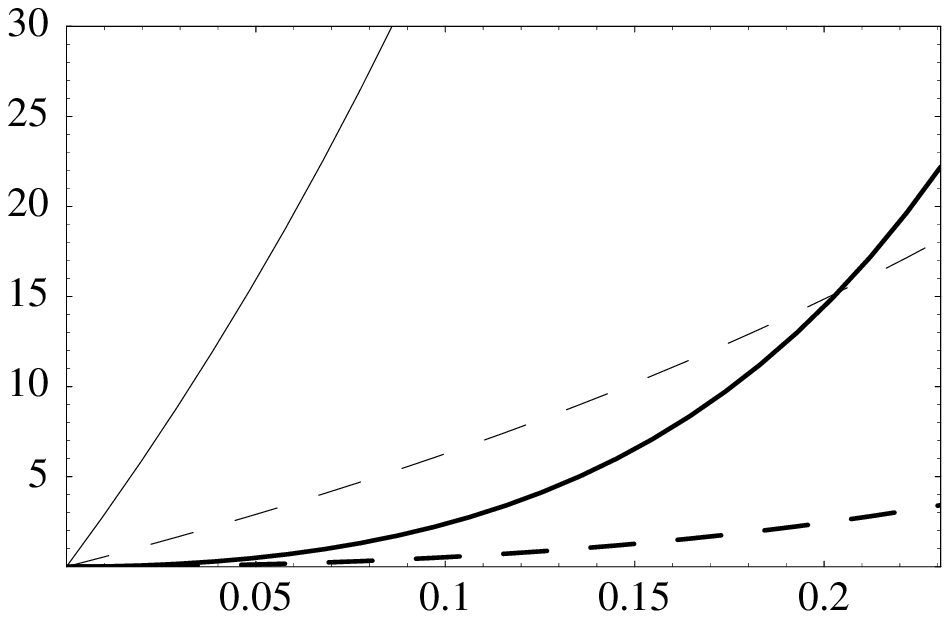,width=0.8\linewidth}}}
\end{picture}
\caption{
\label{errS}
The error in the scalar quantities. The full lines are the 
quadrupole moments, the dashed lines are the amplitudes. The thin lines 
are the leading order corrections, the thick lines are the next-to-leading 
order corrections.}
\end{figure}

\begin{figure}
\setlength{\unitlength}{\linewidth}
\begin{picture}(1,0.6)
\put(0.51,0.05){\makebox(0,0){$\gamma=\epsilon=\delta$}}
\put(0.5,0.3){\makebox(0,0){$e_{A_{\rm T}}^{(0)}$}}
\put(0.73,0.21){\makebox(0,0){$e_{A_{\rm T}}^{(1)}$}}
\put(0.27,0.5){\makebox(0,0){$e_{C_2^{\rm T}}^{(0)}$}}
\put(0.7,0.45){\makebox(0,0){$e_{C_2^{\rm T}}^{(1)}$}}
\put(0.6,0.53){\makebox(0,0){Error in \% }}
\put(0.45,0.33){\makebox(0,0){\epsfig{figure=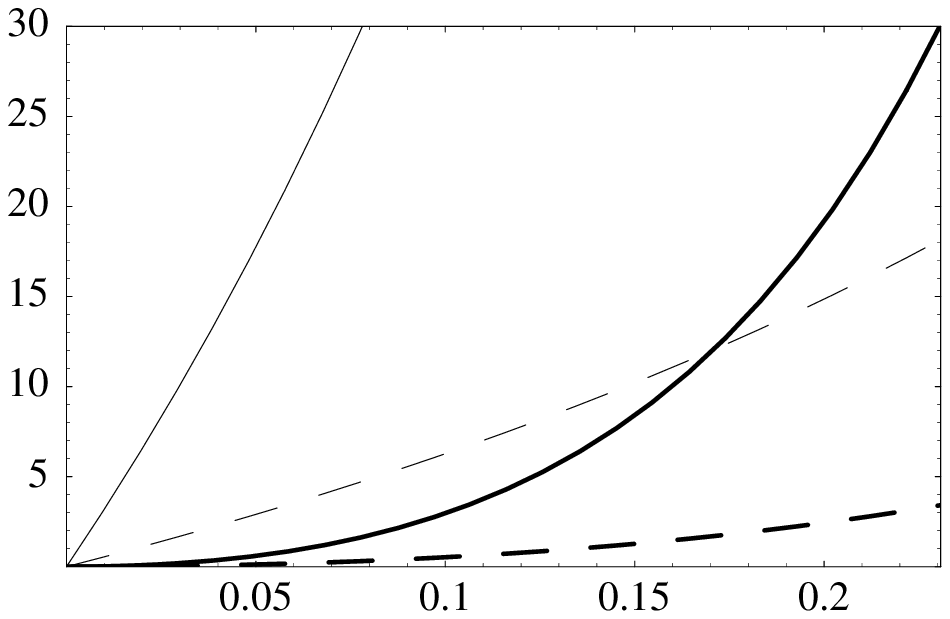,width=0.8\linewidth}}}
\end{picture}
\caption{
\label{errT}
The same as Fig.~\ref{errS}, but for the tensor amplitude and quadrupole 
moment. For the amplitude the error is the same as for the scalar sector,
because $\epsilon = \delta$.}
\end{figure}

So far, except in some of the figures, we did not specify the pivot scale 
$k_0$. We now choose to fix $k_0 \equiv a_* H_*$, i.e. $k_0$ is the mode that 
crosses the horizon at the time $\eta_*$, which is the time when we determine 
the values of the slow-roll parameters, where we fix them once and 
forever. It is easy to show that this amounts to take $\vert \eta _*\vert 
=k_0^{-1}$ in the argument of the logarithm which shows up in
the equations of the previous section. Still $\eta_*$ remains to be fixed, 
which can be done most conveniently by fixing $k_0 r_{\rm lss}$. In the 
following we will show that this choice is of physical relevance, because 
the accuracy of the slow-roll approximation can be improved by a clever 
choice of the pivot scale. We will discuss two cases. The usual convention 
is to choose $k_0 r_{\rm lss} = 1$.  This corresponds to $k_0^{\rm phys} = 
h/(6000 \mbox{\ Mpc})$ today. Below we show that this leads to huge errors 
form the slow-roll approximation.

To improve the precision of the slow-roll approximation we suggest to 
minimize the error in the region of the first acoustic peak, i.e. around 
$l\sim 200$. For this purpose we choose a pivot scale such that
$k_0 r_{\rm lss} \approx 100 e$, which corresponds to a physical wavenumber
$k_0^{\rm phys} = h/(22 \mbox{\ Mpc})$ today. Let us start by analyzing the 
errors for the pivot $k_0 = 1/r_{\rm lss}$.

The errors in the amplitudes and quadrupoles 
for the case of 
power-law inflation, $\epsilon=\delta$, are displayed 
in Figs.~\ref{errS} and \ref{errT}. From these two plots, we can draw 
three conclusions. The first conclusion 
is that the error in the quadrupoles is larger than the error in the 
amplitudes. This confirms the results already obtained in Ref.~\cite{SM}. 
The second conclusion is that it is not possible to obtain an error at 
the $1\%$ level with the leading order, except for very small values of the
slow-roll parameters. The third conclusion is that the accuracy of the 
next-to-leading order for the quadrupoles is better than $1 \%$ if 
$\gamma <0.07$, which corresponds to $0.85 < n_{\rm S}$. Since the slow-roll 
approximation is more accurate for power-law model, it is reasonable to 
expect larger errors for more realistic models. 

\begin{figure}
\setlength{\unitlength}{\linewidth}
\begin{picture}(1,0.6)
\put(0.5,0.05){\makebox(0,0){$l$}}
\put(0.22,0.53){\makebox(0,0){$n_{\rm S} = 0.6$}}
\put(0.4,0.46){\makebox(0,0){$n_{\rm S} = 0.7$}}
\put(0.56,0.36){\makebox(0,0){$n_{\rm S} = 0.8$}}
\put(0.64,0.28){\makebox(0,0){$n_{\rm S} = 0.9$}}
\put(0.74,0.185){\makebox(0,0){$n_{\rm S} = 0.95$}}
\put(0.7,0.53){\makebox(0,0){Error in \% }}
\put(0.45,0.33){\makebox(0,0){\epsfig{figure=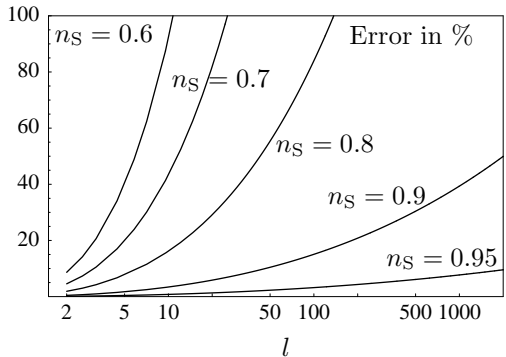,width=0.8\linewidth}}}
\end{picture}
\caption{
\label{errhighl}
The error from the next-to-leading order slow-roll approximation 
in the scalar multipoles $C_l^{\rm S}$ versus $l$ for various 
values of the spectral index $n_{\rm S}$. The approximation is best 
close to todays horizon since, in this figure, we have 
taken $k_0 r_{\rm lss} = 1$, the most common choice.}
\end{figure}

Let us now turn to the slow-roll errors in higher scalar multipole moments. 
Although we cannot obtain the exact result for the high-$l$ multipole moments
without making use of a Boltzmann code, we can nevertheless estimate the 
errors from the slow-roll approximation in this regime. As already discussed 
around Fig.~\ref{errcmb} computing the scalar multipoles from Eq.~(\ref{cls})
is a bad approximation for high $l$. Despite this fact, it is clear that
for a given cosmological model the transfer function is the same for a  
power-law model and its slow-roll approximation. Thus the only difference 
between the power-law and slow-roll multipole moments can arise from the 
convolution of this transfer function with different initial spectra.   
We expect that this difference is small. To put it differently,
$C_l({\rm sr,\ t}) - C_l({\rm sr,\ a})$ is large, whereas
$C_l({\rm sr,\ t}) - C_l({\rm pl,\ t}) \approx C_l({\rm sr,\ a}) - 
C_l({\rm pl,\ a})$, where t (a) denotes the use of the true (approximated) 
transfer function and sr (pl) denotes the initial spectrum.
Thus we use Eqs.~(\ref{C2s}), (\ref{Anpl}), and (\ref{Clsr}) to obtain the 
errors for the scalar multipoles as a function of $l$, (\ref{deferr}), 
which are displayed in Fig.~\ref{errhighl}. It shows that these errors are 
large and increase with $l$ and $|n_{\rm S} -1|$.

\begin{figure}
\setlength{\unitlength}{\linewidth}
\begin{picture}(1,0.6)
\put(0.5,0.05){\makebox(0,0){$l$}}
\put(0.23,0.40){\makebox(0,0){$n_{\rm S} = 0.6$}}
\put(0.18,0.33){\makebox(0,0){$0.7$}}
\put(0.18,0.26){\makebox(0,0){$0.8$}}
\put(0.18,0.19){\makebox(0,0){$0.9$}}
\thinlines
\put(0.17,0.145){\line(3,1){0.3}}
\put(0.57,0.245){\makebox(0,0){$n_{\rm S} = 0.95$}}
\put(0.7,0.53){\makebox(0,0){Error in \% }}
\put(0.45,0.33)
{\makebox(0,0){\epsfig{figure=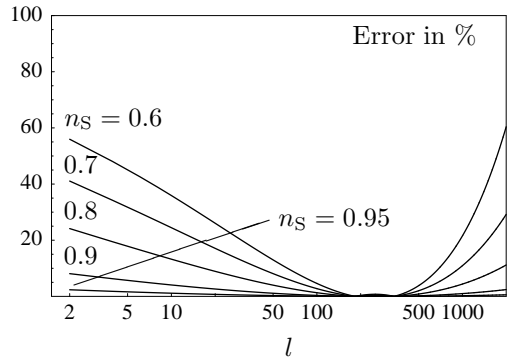,width=0.8\linewidth}}}
\end{picture}
\caption{
\label{errhighl_impr}
The same as figure \ref{errhighl}, but with $k_0 r_{\rm lss} = 100 e$. This 
optimizes the accuracy in the region of the first acoustic peak.}
\end{figure}

The reason for the large errors in the multipole moments are the large errors
in the spectral indices. This can be understood from the relations 
between the errors:
\begin{eqnarray}
\label{e0}
{e^{(0)}_{C_l^{\rm S}}\over 100} &=& \left| {g_l(1)\over g_l(n_{\rm S})}
\left(1 + {e^{(0)}_{A_{\rm S}}\over 100}\right) - 1\right| \ , \\
{e^{(1)}_{C_l^{\rm S}}\over 100} &=& \left| {g_l(1)\over g_l(n_{\rm S})}
\left(1 + {e^{(1)}_{A_{\rm S}}\over 100}\right) + 
{g_l'(1) (n_{\rm S}-1) \over g_l(n_{\rm S})} 
\right. \nonumber \\ 
\label{e1}
& & \left. \times
\left(1 + {e^{(0)}_{A_{\rm S}}\over 100}\right)
\left(1 - {e^{(1)}_{n_{\rm S}-1}\over 100}\right) - 1\right| \ ,
\end{eqnarray}
where the prime denotes a derivative with respect to $n_{\rm S}$.
The signs in front of the errors in the amplitude and in the spectral 
index are model dependent. For $\delta = \epsilon$ the error in the 
amplitude has always positive sign, the error in the spectral index
has a negative sign. For small values of $l$ we may expand 
\begin{equation}
\label{expg}
g_l(n_{\rm S}) = g_l(1) + g_l'(1)(n_{\rm S} -1) + {\cal O}[(n_{\rm S} -1)^2]
\end{equation}
in Eqs.~(\ref{e0}) and (\ref{e1}). Keeping only terms linear in $n_{\rm S} -1$
and terms linear in the errors we find 
\begin{equation}
\label{decomperror}
e_{C^{\rm S}_l}^{(i)} \approx e_{A_{\rm S}}^{(i)} + 
(D_l -\ln k_0r_{\rm lss})(1 - n_{\rm S}) e_{n_{\rm S} -1}^{(i)} \ ,  
\end{equation}
where $D_l$ has been introduced in Eq.~(\ref{Dl}). The error in the quadrupole 
moment is now easily understood from the last equation. As claimed above,
the large error in the spectral index is responsible for the large error in
the quadrupole moment. The contribution from the error in the spectral index  
always dominates. It is obvious from Fig.~\ref{errhighl} that the error 
increases
with $l$ and with $|n_{\rm S} -1|$. Equation (\ref{decomperror}) underestimates
the error for large $l$ and $|n_{\rm S} -1|$, due to the breakdown of the
expansion (\ref{expg}). 

These large errors, displayed in Fig.~\ref{errhighl}, can be shifted to 
different multipoles by a change of the pivot scale $k_0$. Therefore, one may  
hide part of the error from the slow-roll approximation in the cosmic 
variance. Inspection of Eq.~(\ref{decomperror}) suggests that the error is 
minimized for a given multipole if $D_l = \ln(k_0 r_{\rm lss})$. For large $l$ 
\begin{equation}
k_0 \approx \frac{e}{2}\frac{l}{r_{\rm lss}}, \quad \mbox{or} \quad  
k_0^{\rm phys} \approx \frac e4 l H_0,  
\end{equation}
where $H_0$ is todays Hubble rate. We would like to know the position and 
hight of the first acoustic peak very accurate, which suggests to choose 
$l_{\rm minimal\ error} \approx 200$. This gives $k_0 = 1/22 h^{-1}$ Mpc 
$\approx 1/31$ Mpc for $h = 0.71$, the HST key project final value. 
Figure \ref{errhighl_impr} shows the errors at the next-to-leading order for 
various values of the scalar index with the new choice for the pivot scale.
It can be seen clearly that the errors are highly suppressed around 
$l \approx 200$, as expected, but increase at lower and higher multipoles. 
The tiny bump between $l \approx 200$ and $l \approx 500$ is due to the fact 
that we plot the absolute value, in this region the error changes its sign. 
The new choice of the pivot scale allows to predict the multipoles in the 
range $2\leq l \leq 2000$ for $n_{\rm S} = 0.9$ better than $10\%$, which was 
not possible with the pivot scale chosen previously. Nevertheless, the 
precision is not good enough to reach the $1\%$ accuracy level (the error is 
$2.4\%$ at $l = 2000$). In order to do so it is necessary to have 
$n_{\rm S} > 0.93$ or $\gamma < 0.032$.
 
\begin{figure}
\setlength{\unitlength}{\linewidth}
\begin{picture}(1,0.6)
\put(0.51,0.05){\makebox(0,0){$\gamma=\epsilon=\delta$}}
\put(0.47,0.35){\makebox(0,0){$e_R^{(0)}$}}
\put(0.74,0.28){\makebox(0,0){$e_R^{(1)}$}}
\put(0.6,0.53){\makebox(0,0){Error in \% }}
\put(0.45,0.33){\makebox(0,0){\epsfig{figure=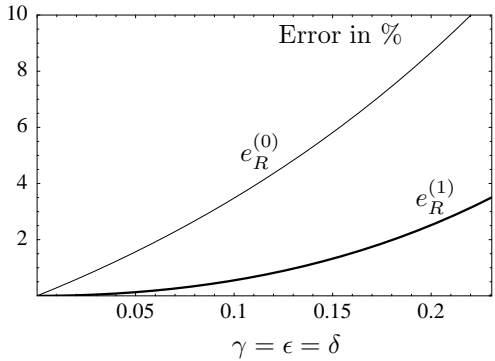,width=0.8\linewidth}}}
\end{picture}
\caption{
\label{errR}
The error in the tensor to scalar ratio in leading (thin line) 
and next-to-leading order (thick line).}
\end{figure}

The error in the $T/S$ ratio is displayed in Fig.~\ref{errR}. In our special 
situation the pivot scale does not enter $R$, because
$\epsilon = \delta$ [see Eq.~(\ref{Rsr})]. We see that the error 
in $R$ is less important than for the 
amplitudes and/or the quadrupoles. Therefore this suggests 
to use $R$ to test the single scalar field/slow-roll paradigm. However, 
it is clear that any violation of the consistency check by the 
forthcoming data should be interpreted as a failure of this 
paradigm but not as the failure of inflation itself. In a 
more general situation where $\epsilon \neq \delta $, the choice 
of $k_0$ does affect the error in $R$. Since we do not have an exact 
solution for scalar and tensors modes such that $\epsilon \neq \delta $ at 
our disposal, it is difficult to predict the corresponding effect. 
\par
The errors from the slow-roll approximation displayed in 
Figs.~\ref{errS}, \ref{errT}, \ref{errhighl}, 
and \ref{errhighl_impr} should be compared 
with the cosmic variance. The cosmic variance is the variance of the 
best unbiased estimator for the multipole moments \cite{GM,GanM}: 
${\cal E}_{\rm Best}(C_l)=1/(2l+1)\sum_{m=-l}^{m=l}a_{lm}a_{lm}^*$, 
where we have expanded the temperature fluctuations over the basis of
spherical harmonics, $\Delta (\vec{e})=\sum _{lm}a_{lm}Y_{lm}(\vec{e})$. This 
expression is valid only if the $a_{lm}$'s satisfy a Gaussian or a mildly 
non-Gaussian statistics. The corresponding error can be written as
\begin{equation}
\label{cosvar}
e_{C_l}^{\rm (cv)}\equiv   
\frac{\sigma _{{\cal E}_{\rm Best}(C_l)}}{C_l}\times 100 \%
= \sqrt{\frac{2}{2l+1}} \times100\%\ .
\end{equation}
Over the whole range of the spectrum that will be measured by a mission like 
Planck the cosmic variance is larger than $2\% (l \approx 2000)$ and is 
$\approx 7\%$ at the first acoustic peak $(l \approx 200)$.

It is possible to reduce the cosmic variance by binning 
several multipoles together at the expense of decreasing the precision 
on the location of the multipoles. Therefore, we define an averaged 
multipole, $\bar C_{l}$, on the range $[l-L,l+L]$ by 
\begin{equation}
\label{defavemulti}
\bar C_l \equiv \frac{1}{2 L + 1}
\sum _{j=l-L}^{l+L}j(j+1)C_j\ .
\end{equation}
The central value $l$ of each interval must 
be separated by $2 L+1$. For an incomplete sky coverage $L$ is 
restricted by the form of the basis used to expand $\Delta (\vec{e})$. 
Below we do not 
take this issue into account and assume that the full sky is covered. 
In order to calculate the cosmic variance associated with the binned 
multipole, we define the estimator
\begin{equation}
\label{estibin}
{\cal E}(\bar C_l)\equiv \frac{1}{2L+1}
\sum_{j=l-L}^{l+L}\frac{j(j+1)}{2j+1}
\sum _{m=-j}^j a_{jm}a_{jm}^*.
\end{equation}
This estimator is clearly unbiased, 
$\langle {\cal E}(\bar C_l)\rangle 
=\bar C_l$, and it is very likely that this is also the best one 
although a rigorous proof is not presented here. Its variance can be 
expressed as 
\begin{equation}
\sigma _{{\cal E}(\bar C_l)}^2= {1\over (2 L +1)^2}
\sum_{j=l-L}^{l+L} j^2(j+1)^2 {2 C_j^2\over 2j+1}\ .
\end{equation}
Using that $l(l+1) C_l \approx$ const, if $L$ is not too big, we arrive at 
\begin{equation}
\label{cosvarbin}
e_{\bar C_l}^{\rm (cv)}\equiv  
\frac{\sigma_{{\cal E}(\bar C_l)}}{\bar C_l}\times  100\%
\approx \frac{1}{2L+1} \sqrt{\sum_{j=l-L}^{l+L}
\frac{2}{2j+1}} \times 100 \%\ . 
\end{equation}
For $L=0$ this error reduces to the known expression for the cosmic variance
(\ref{cosvar}). In this case it is independent of the spectral index, whereas
for $L\neq 0$ this is true as long as $l(l+1) C_l$ is approximately constant 
within the range $[l-L,l+L]$. With the same approximation we find that the
slow-roll error in $\bar C_l$ is just $e^{(i)}_{\bar C_l} \approx 
e^{(i)}_{C_l}$. 

The cosmic variance for different multipoles and binning is displayed in 
table \ref{tab}. For comparison we give the errors in the multipole moments 
from the slow-roll approximation at leading and next-to-leading order
for the pivot scale corresponding to the present horizon and we also present 
the errors at the next-to-leading order for the pivot scale corresponding 
to the scale of the first acoustic peak. We present the results for two 
values of the spectral index, corresponding to $1(2)\sigma$ errors in the COBE
measurement. 

For $k_0=r_{\rm lss}^{-1}$, the error from the slow-roll approximation in 
leading order
dominates over the cosmic variance already at $l=10$ for $n_{\rm S} = 0.9$ and
$L=0$. For the next-to-leading order and $n_{\rm S} = 0.9$ the slow-roll 
error dominates over the cosmic variance at $l=100$, for any binning of 
multipoles. In the case of $n_{\rm S} = 0.6$ only the error in the quadrupole 
from the slow-roll approximation at next-to-leading order is smaller than the 
cosmic variance. Only for $|n_{\rm S} - 1| < 10^{-2}$ the error from 
the next-to-leading order is below $1\%$, which is the order of magnitude of
the cosmic variance in the Silk damping regime of the spectrum.  

For $k_0 = 1/22h^{-1}\mbox{Mpc}$, the situation is improved. The 
slow-roll error and the cosmic variance are of the same order of magnitude 
up to $l=1000$ ($n_{\rm S}=0.9$). However, this is not sufficient to 
hide the slow-roll error in the cosmic variance for the whole spectrum. 
For $n_{\rm S}=0.6$, the slow-roll error exceeds the cosmic variance at 
small and large scales and is hidden in the cosmic variance only for a narrow 
range of scales around $k_0$. Let us point out that values of the tilt
of the order of $0.1$ and larger are a realistic possiblity, as is clearly 
seen from the maximum likelyhood fits to the recent BOOMERanG 
and MAXIMA-1 data.

We conclude that for a general model of inflation only numerical 
mode-by-mode integration can presently provide predictions for the 
CMBR with less than $1\%$ error, unless all slow-roll parameters are less
than $10^{-2}$. To give an example let us consider the case of
chaotic inflation with a potential $V \propto \varphi^p$, for which the 
slow-roll parameters are $\epsilon \approx p/200, \delta \approx (p-2)/200$,
and $\xi \approx (p-2)(p-4)/(200 p)$, giving $n_{\rm S}^{\rm (sr)} \approx
1 - (p+2)/100$. Thus for $p=2(4)$ the errors from the slow-roll approximation 
at next-to-leading order are of the same order as the cosmic variance
for large $l$, since $\epsilon \approx 0.01(0.02)$ and $\delta \approx
0 (0.01)$, which corresponds to $n_{\rm S}^{\rm (sr)} \approx 0.96 (0.94)$.  
However, already for $p=6$, the slow-roll approximation in the next-to-leading 
order leads to errors that exceed the cosmic variance at high multipoles
($\epsilon \approx 0.03, \delta \approx 0.02, \xi \approx 0.007$, thus 
$n_{\rm S}^{\rm (sr)} \approx 0.92$).  

Forthcoming high precision missions, especially the MAP and Planck satellites,
will only be limited by the cosmic variance up to $l \approx 1000$ and 
$l \approx 2000$ respectively. Therefore predictions from inflationary models 
should be made such that the slow-roll error does not exceed the cosmic 
variance. We have shown that there are slow-roll models which cannot meet
this requirement.

Another implication is that the large errors in the predicted multipoles 
render all attempts to reconstruct the inflationary potential difficult.
The reason for this is that reconstruction usually assumes that the 
primordial spectrum, instead of the multipoles, is measured to a high 
precision. We have shown in this work, 
that the errors in the prediction of the multipoles are easily an order of 
magnitude larger. A first attempt to go directly from inflation to the 
calculation of the multipole moments has been put forward by Grivell and 
Liddle \cite{GL2} recently. In our opinion a purely numerical approach to 
this fundamental issue is not fully satisfactory --- better analytic methods 
are needed. 

\begin{table}
\begin{tabular}{c c c c c c c}
$n_{\rm S}$ & $L$ & $l$ & $e_{\bar C_l}^{\rm (cv)}$ & 
$e_{\bar C_l}^{(0)}$ & $e_{\bar C_l}^{(1)}$ &  $e_{\bar C_l}^{\rm impr}$ \\ 
\hline 
$0.9$ & $0$ &     $2$ & $63\%$  & $15\%$ & $0.4\%$ & $8.1\%$  \\
      &     &    $10$ & $31\%$  & $34\%$ & $3.4\%$ & $4.1\%$  \\
      &     &   $100$ & $10\%$  & $68\%$ & $15\%$  & $0.35\%$ \\
      &     &  $1000$ & $3.2\%$ & $-$ & $39\%$     & $1.0\%$  \\
      & $2$ &    $10$ & $14\%$  & $34\%$ & $3.4\%$ & $4.1\%$  \\ 
      &     &   $100$ & $4.5\%$ & $68\%$ & $15\%$  & $0.35\%$ \\
      &     &  $1000$ & $1.4\%$ & $-$ & $39\%$     & $1.0\%$  \\
      & $4$ &    $10$ & $11\%$  & $34\%$ & $3.4\%$ & $4.1\%$  \\
      &     &   $100$ & $3.3\%$ & $68\%$ & $15\%$  & $0.35\%$ \\
      &     &  $1000$ & $1.1\%$ & $-$ & $39\%$     & $1.0\%$  \\
$0.6$ & $0$ &     $2$ & $63\%$  & $73\%$ & $8.6\%$ & $56\%$  \\
      &     &    $10$ & $31\%$  & $-$ & $92\%$     & $36\%$  \\
      &     &   $100$ & $10\%$  & $-$ & $-$        & $4.6\%$ \\
      &     &  $1000$ & $3.2\%$ & $-$ & $-$        & $22\%$  \\
      & $2$ &    $10$ & $14\%$  & $-$ & $92\%$     & $36\%$  \\
      &     &   $100$ & $4.5\%$ & $-$ & $-$        & $4.6\%$ \\
      &     &  $1000$ & $1.4\%$ & $-$ & $-$        & $22\%$  \\
      & $4$ &    $10$ & $11\%$  & $-$ & $92\%$     & $36\%$  \\
      &     &   $100$ & $3.3\%$ & $-$ & $-$        & $4.6\%$ \\
      &     &  $1000$ & $1.1\%$ & $-$ & $-$        & $22\%$ 
\end{tabular}
\caption{
\label{tab}
Comparison of errors from the cosmic variance and the slow-roll
approximation in leading and next-to-leading order with 
$k_0 r_{\rm lss} = 1$ and in next-to-leading order with 
$k_0 r_{\rm lss} = 100 e$, for various values of
the spectral index $n_{\rm S}$, the bin width $L$, and the multipole $l$. 
The ``$-$'' denotes errors that exceed $100\%$.}
\end{table} 

\acknowledgements

We would like to thank A.~R.\ Liddle, V.~F.\ Mukhanov, and  V.\ Shani 
for valuable discussions and/or comments and an anonymous referee for 
suggesting to change the pivot scale. D.~S.\ thanks the Austrian 
Academy of Sciences for financial support. J.~M.\ thanks Robert Brandenberger 
and the High Energy Group of Brown University (Providence, USA) and the 
Institute f\"ur Theoretische Physik (Vienna, Austria) for warm hospitality.


\begin{thebibliography}{99}
\bibitem{Boomerang} P.~de Bernadis et al., Nature {\bf 404}, 955 (2000);
         A.~E.~Lange et al., astro-ph/0005004 (2000); Boomerang
         {\tt http://oberon.roma1.infn.it/boomerang/}.
\bibitem{Maxima}S.~Hanany et al., astro-ph/0005123 (2000);
         A.~Balbi et al., astro-ph/0005124 (2000); \\
         Maxima-1 {\tt http://cfpa.berkeley.edu/group/cmb/}.
\bibitem{MAP+Planck} MAP (Microwave Anisotropy Probe)\\ 
         {\tt http://map.gsfc.nasa.gov/};\\ 
         Planck {\tt http://astro.estec.esa.nl/SA-general/\\
         Projects/Planck/}.
\bibitem{inflation} A.~Guth, Phys. Rev. D {\bf 23}, 347 (1981);
         A.~Linde, Phys. Lett. B {\bf 108}, 389 (1982);
         A.~Albrecht and P.~J.~Steinhardt, Phys. Rev. Lett. {\bf 48}, 1220 
         (1982); A.~Linde, Phys. Lett. B {\bf 129}, 177 (1983).
\bibitem{tensor}A.~A.~Starobinsky, Pis'ma Zh. Eksp. Teor. Fiz.
         {\bf 30}, 719 (1979) [JETP Lett. {\bf 30}, 682 (1979)].
\bibitem{scalar}V.~Mukhanov and G.~Chibisov, JETP Lett. {\bf 33}, 532
         (1981); S.~Hawking, Phys. Lett. {\bf 115B}, 295 (1982);
         A.~A.\ Starobinsky, Phys. Lett. {\bf 117B}, 175 (1982);
         J.~M.~Bardeen, P.~J.~Steinhardt, and M.~S.~Turner, Phys. Rev. D 
         {\bf 28}, 679 (1983).
\bibitem{GP}A.~Guth and S.~Y.~Pi, Phys. Rev. Lett. {\bf 49}, 1110 (1982).
\bibitem{MFB}V.~F.~Mukhanov, H.~A.~Feldman and R.~H.~Brandenberger, 
         Phys. Rep. {\bf 215}, 203 (1992).
\bibitem{Grivell}I.~J.~Grivell and A.~R.~Liddle, Phys. Rev. D {\bf 54}, 7191
         (1996).
\bibitem{SL}E.~D.~Stewart and D.~H.~Lyth, Phys. Lett. B {\bf 302}, 171 
         (1993).
\bibitem{WMS}L.~Wang, V.~F.~Mukhanov and P.~J.~Steinhardt, 
         Phys. Lett. B {\bf 414}, 18 (1997).
\bibitem{Lea}J.~E.~Lidsey et al., Rev. Mod. Phys. {\bf 69}, 373 (1997).
\bibitem{Coea}E.~J.~Copeland et al., Phys. Rev. D {\bf 49}, 1840 (1994).
\bibitem{KV}E. W.~Kolb and S.~L.~Vadas, Phys. Rev. D {\bf 50}, 2479 (1994).
\bibitem{Cea2}E.~J.~Copeland et al., Phys. Rev. D {\bf 58}, 043002 (1998).
\bibitem{StaroExact}A.~A.~Starobinsky, talk at DARC, Meudon, December 11, 
         1998.
\bibitem{B}J.~A.~Bardeen, Phys. Rev. D {\bf 22}, 1882 (1980).
\bibitem{Grigw}L.~P.~Grishchuk, Zh. Eksp. Teor. Fiz. {\bf 64}, 825 (1974)
         [Sov. Phys. JETP {\bf 40}, 409 (1974)].
\bibitem{M}V.~F.~Mukhanov, Zh. Eksp. Teor. Fiz. {\bf 94}, 1 (1988) [Sov.
         Phys. JETP {\bf 68}, 1297 (1988)].
\bibitem{Griden} L.~P.~Grishchuk, Phys. Rev. D {\bf 50}, 7154 (1994).
\bibitem{MS} J.~Martin and D.~J.~Schwarz, Phys. Rev. D {\bf 57}, 3302 (1998).
\bibitem{COBE1}G.~Smoot et al., Astrophys. J. {\bf 396}, L1 (1992);
         C.~L.~Bennett et al., Astrophys. J. {\bf 464}, L1 (1996);
         K.~M.~Gorski et al., Astrophys. J. {\bf 464}, L11 (1996).
\bibitem{SW} R.~K.~Sachs and A.~M.~Wolfe, Astrophys. J. {\bf 147}, 73 (1967).
\bibitem{FIRAS}D.~J.~Fixsen et al., Astrophys. J. {\bf 473}, 576 (1996). 
\bibitem{GM} L.~P.~Grishchuk and J.~Martin, Phys. Rev. D {\bf 56}, 1924 (1997).
\bibitem{P82} P.~J.~E.~Peebles, Ap. J. {\bf 263}, L1 (1982).
         J.~R.~Bond and G.~Efstathiou, Ap. J. {\bf 285}, L45 (1984).
\bibitem{Staro} A.~A.~Starobinsky, Sov. Astron. Lett. {\bf 11}, 133 (1985).
\bibitem{Dea} R.~L.~Davis et al., Phys. Rev. Lett. {\bf 69}, 1856 (1992).
         D.~S.\ Salopek, Phys. Rev. Lett. {\bf 69}, 3602 (1992).
         A.~R.~Liddle and D.~H.~Lyth, Phys. Lett. B {\bf 291}, 391 (1992).
\bibitem{SS}T.~Souradeep and V.~Sahni, Mod. Phys. Lett. {\bf 7}, 3541 (1992).
\bibitem{Cea} R.~Crittenden et al., Phys. Rev. Lett. {\bf 71}, 324 (1993).
\bibitem{St} P.~J.~Steinhardt, Int. J. Mod. Phys. A {\bf 10}, 1091 (1995).
\bibitem{transfer} J.~Bardeen et al., Astrophys. J. {\bf 304}, 15 (1986).
\bibitem{SZ}U.~Seljak and M.~Zaldarriaga, Astrophys. J. {\bf 469}, 
         437 (1996); CMBFAST Website: \\
         {\tt http://www.sns.ias.edu/\~{}matiasz/CMBFAST/ \\
         cmbfast.html}.
\bibitem{BN} E.~Bunn and M.~White, Astrophys. J. {\bf 480}, 6 (1997).
\bibitem{L}D. H.~Lyth, Phys. Rev. D {\bf 31}, 1792 (1985).
\bibitem{SM}D.~J.~Schwarz and J.~Martin, astro-ph/9805313 
         published in: {\em Current Topics in Mathematical Cosmology}, 
         Eds. M. Rainer and H.-J. Schmid (World Scientific PC, Singapore, 
         1998) pp. 65.
\bibitem{GanM} A.~Gangui and J.~Martin, astro-ph/0001361.
\bibitem{GL2}I. J.~Grivell and A.~R.~Liddle, Phys. Rev. D {\bf 61}, 081301
         (2000).
\end{thebibliography}
\end{document}